\documentclass[]{elsart}

\usepackage{natbib}
\usepackage{graphics}
\usepackage{graphicx}
\usepackage{epsfig}
\usepackage{amssymb}

\begin{document}

\begin{frontmatter}

\title{Endochronic theory, non-linear kinematic
hardening rule and generalized plasticity: a new interpretation
based on generalized normality assumption}

\author[a]{Silvano Erlicher\corauthref{cor1}}
\ead{erlicher@lami.enpc.fr}
\author[a,b]{, Nelly Point}
\corauth[cor1]{Corresponding author. Tel: +33 1 64 15 37 80, Fax:
+33 1 64 15 37 41}

\address[a]{ Laboratoire d'Analyse des Mat\'{e}riaux
  et Identification LAMI (ENPC/LCPC-Institut Navier),
6 et 8 av. B. Pascal, Cit\'{e} Descartes, Champs-sur-Marne, 77455
Marne-la-Vall\'{e}e, Cedex 2, France}

\address[b]{ Conservatoire National des Arts et M\'{e}tiers CNAM,
Sp\'{e}cialit\'{e} Math\'{e}matiques (442), 292 rue Saint-Martin,
75141 Paris, Cedex 03, France}

\begin{abstract}
Short title: Non-classical plasticity theories and generalized
normality.
\newline
\newline
A simple way to define the flow rules of plasticity models is the
assumption of generalized normality associated with a suitable
pseudo-potential function. This approach, however, is not usually
employed to formulate endochronic theory and non-linear kinematic
(NLK) hardening rules as well as generalized plasticity models. In
this paper, generalized normality is used to give a new
formulation of these classes of models. As a result, a suited
pseudo-potential is introduced for endochronic models and a
non-standard description of NLK hardening and generalized
plasticity models is also provided. This new formulation allows
for an effective investigation of the relationships between these
three classes of plasticity models.
\end{abstract}


\end{frontmatter}


\section{Introduction}

The models proposed so far in the literature to describe the rate
independent inelastic behavior of real materials subjected to
monotonic or cyclic loading conditions can be essentially
classified into two main families: (i) models where the present
state depends on the present value of observable variables (total
strain, temperature) and of suitable internal variables; (ii)
models, indicated here as \emph{hereditary}, that require the
knowledge of the whole past history of observable variables.

The first group encompasses, for instance, the classical models of
Prandtl-Reuss and Prager (see e.g. \citet{Lemaitre90engl}) and the
NLK hardening model of \cite{Armstrong66}, in its original form as
well as in the modified versions recently proposed by
\citet{Chaboche91} and \citet{Ohno93} in order to improve the
ratchetting modelling. For these models, the well known notions of
\emph{elastic domain} and \emph{loading} (or \emph{ yielding})
\emph{surface} apply. \emph{Associativity} and
\emph{non-associativity} of the plastic strain flow rule are also
well-established concepts, as well as the assumption of
generalized associativity (or \emph{generalized normality}),
relating all internal variable flow directions to a given loading
surface \citep{Halphen75} \citep{JirasekBazant02}. Using the
language of convex analysis \citep{Rockafellar69}, generalized
normality entails that the flows of all internal variables belong
to the \emph{subdifferential set} of a given scalar non-negative
function called \emph{pseudo-potential} \citep{Moreau70}
\citep{Fremond2002}.

Among internal variable theories, \emph{generalized plasticity}
deserves special attention. A first important step for its
formulation was the idea, suggested by \cite{Eisenberg71}, of a
plasticity model where, despite classical plasticity, loading and
yielding surfaces are $\emph{not}$ coincident. Then, starting from
an axiomatic approach to describe inelastic behavior of materials,
Lubliner proposed some simple generalized plasticity models, able
to represent some observed experimental behavior of metals
(Lubliner, 1974, 1980, 1984) \citep{Lubliner93}. More recently,
generalized plasticity has been used for describing the shape
memory alloy behavior \citep{Lubliner96}.

\emph{Endochronic} models \citep{Valanis71} and \emph{Bouc-Wen}
type models \citep{Bouc71} \citep{Wen76} are two important
examples of hereditary models. Endochronic theory has been
developed during the seventies and used for modelling the plastic
behavior of metals (see, for instance, \citet{Valanis71},
\citet{Valanis75}) and the inelastic behavior of concrete and
soils (among others, \cite{BazantKrizek76}, \cite{Bazant76}). The
endochronic stress evolution rule depends on the so-called
\emph{intrinsic time} and is formulated by a convolution integral
between the strain tensor and a scalar function of the intrinsic
time called \emph{memory kernel}. When the kernel is an
exponential function, an incremental form of endochronic flow
rules exists, which is commonly used in standard analyses and
applications.

Models of Bouc-Wen type are widely employed for modelling the
cyclic behavior of structures in seismic engineering
\citep{Baber81} \citep{Sivaselvan2000} and for representing
hysteresis of magneto-rheological dampers in semi-active control
applications \citep{Sain97} \citep{Jansen2000}. The strict
relationship between endochronic and Bouc-Wen type models has been
mentioned several times in the literature (see, among others,
\cite{Karray89} and \cite{Casciati89}). Recently,
\citet{Erlicher04} showed that the fundamental element of this
relationship is the choice of an appropriate intrinsic time.

Endochronic theory and classical internal variable theory have
been compared by using several approaches: \cite{Bazant78}
observed that for endochronic theory the notion of loading surface
can still be introduced, but it looses its physical meaning;
\cite{Valanis80} and \cite{Watanabe86} proved that a NLK hardening
model can be derived from an endochronic model by imposing a
special intrinsic time definition. Moreover, a comparative study
between NLK hardening and generalized plasticity models has been
presented by \cite{Auricchio95}. A tight relationship between
endochronic theory and generalized plasticity is also expected to
exist, but, by the authors' knowledge, no analysis on this subject
exists. More generally, there is a lack of unified theoretical
framework, on which formal comparisons between these plasticity
theories could be based. The main goal of this paper is the
formulation of this theoretical framework using the classical
notion of \emph{generalized normality} \citep{Moreau70}
\citep{Halphen75}. As a result, a new formulation of endochronic
and NLK hardening models as well as generalized plasticity models
is suggested and is used to investigate the relationships between
them.

The paper is organized as follows: in the first section, the
standard theoretical framework of thermomechanics is briefly
recalled, with reference to the notions of pseudo-potential and
generalized normality as well as Legendre-Fenchel transform and
dual pseudo-potential. In the following sections, several
plasticity models are presented and are shown to fulfil the
generalized normality assumption. The Prandtl-Reuss and
endochronic models are considered first, in both standard and
multi-layer formulations. Then, NLK hardening model and
generalized plasticity follow. The discussion is limited to
initially isotropic materials, whose plastic behavior is governed
by the second invariant of the deviatoric stress, $J_{2}$, known
as von Mises or $J_{2}$ materials. No stability analysis is
provided, as it is beyond the purposes of this contribution.

\section{General thermodynamic framework}

Under the assumption of infinitesimal transformations, the
classical expression of the local form of the first and second
principle of thermodynamics can be written as follows:
\begin{equation}
T\dot{s}=-\dot{\Psi}-s\dot{T}-div\left( \mathbf{q}\right)
+r+\mbox{\boldmath $\sigma$}:\mbox{\boldmath $\dot{\varepsilon}$}
\label{1princ}
\end{equation}
\begin{equation}
\Phi _{s}\left( t\right) =\dot{s}+div\left(
\frac{\mathbf{q}}{T}\right) - \frac{r}{T}\geq 0  \label{2ndprinc0}
\end{equation}
where the superposed dot indicates the time derivative; $s$ is the
entropy density per unit volume, $\mathbf{q}$ is the vector of the
flowing out heat flux, $T$ is the absolute temperature, $r$ is the
rate of heat received by the unit volume of the system from the
exterior; $\mbox{\boldmath $\sigma$}$ is the second order
symmetric Cauchy stress tensor; $\mbox{\boldmath $\varepsilon$}$
is the tensor of small total strains; $\Phi _{s}\left( t\right) $
is the rate of interior entropy production. In the vector space of
all second order tensors, the Euclidean scalar product $:$ is
defined by $\mathbf{x:y}=x_{ij} y_{ij}$; the vector subspace of
second order symmetric tensors is denoted by $ \mathbb{S}^{2}$.
The Helmholtz free energy density per unit volume is a state
function defined as
\begin{equation}
\Psi =\Psi \left( \mbox{\boldmath $\varepsilon$},T,\mbox{\boldmath
$\chi$}_{1},...,\mbox{\boldmath $\chi$}_{N}\right) =\Psi \left(
\mathbf{v}\right)   \label{Potenz}
\end{equation}
where $\mbox{\boldmath $\chi$}_{1},...,\mbox{\boldmath
$\chi$}_{N}$, are the tensorial and/or scalar internal variables,
related to the non-elastic evolution and $ \mathbf{v=}\left\{
\mbox{\boldmath $\varepsilon$},T,\mbox{\boldmath
$\chi$}_{1},...,\mbox{\boldmath $\chi$}_{N}\right\} $ is the
vector containing all the state variables, namely the total strain
tensor, the temperature and the internal variables.

For isothermal conditions, the use of Eq. ( \ref{1princ}) in the
inequality (\ref{2ndprinc0}) leads to
\begin{equation}
\Phi_{m}\left( t\right) :=T\Phi _{s}\left( t\right)
=T\dot{s}=\mbox{\boldmath $\sigma$}:\mbox{\boldmath
$\dot{\varepsilon}$} -\dot{\Psi}\geq 0 \label{2ndprinc}
\end{equation}
which states that the \emph{intrinsic }or \emph{mechanical\
dissipation} $ \Phi _{m}$ (rate of energy per unit volume) must be
non-negative. The \emph{ non-dissipative} thermodynamic forces are
defined as functions of the free energy density $\Psi $ (see,
among others, \cite{Fremond2002})
\begin{equation}
\begin{array}{c}
\mbox{\boldmath $\sigma$}^{nd}:=\frac{\partial \Psi }{\partial
\mbox{\scriptsize \boldmath $\varepsilon$}}, \ \ \ \
\mbox{\boldmath $\tau$}_{i}^{nd}:=\frac{\partial \Psi }{\partial
\mbox{\scriptsize\boldmath $\chi$}_{i}}\ \ \ \Longleftrightarrow \
\ \ \ \ \mathbf{q}^{nd}=\frac{\partial \Psi }{\partial \mathbf{v}}
\end{array}
\label{fnondiss}
\end{equation}
The non-dissipative stress $\mbox{\boldmath $\sigma$}^{nd}$ is
associated with the observable variable $\mbox{\boldmath
$\varepsilon$}$, while $\mbox{\boldmath $\tau$}_{i}^{nd}$ are
associated with the internal variables $\mbox{\boldmath
$\chi$}_{i}$. All non-dissipative forces can be collected in
$\mathbf{q}^{nd}=\left(\mbox{\boldmath
$\sigma$}^{nd},\mbox{\boldmath
$\tau$}_{1}^{nd},...,\mbox{\boldmath $\tau$}_{N}^{nd}\right) $.
Hence, by substituting (\ref{fnondiss}) into (\ref{2ndprinc}), one
obtains
\begin{equation}
\Phi _{m}\left( t\right) =\left( \mbox{\boldmath
$\sigma$}-\mbox{\boldmath $\sigma$} ^{nd}\right) :\mbox{\boldmath
$\dot{\varepsilon}$}-\sum_{i=1}^{N}\mbox{\boldmath $\tau$}
_{i}^{nd}\cdot \mbox{\boldmath $\dot{\chi}$}_{i}=\mbox{\boldmath
$\sigma$}:\mathbf{\mbox{\boldmath
$\dot{\varepsilon}$}-q}^{nd}\cdot \mathbf{\dot{v}}\geq 0
\label{2princ2}
\end{equation}
where $\mathbf{\dot{v}}$ is the vector of the fluxes, belonging to
a suitable vector space $\mathbb{V}$. The vector spaces considered
in this paper are isomorph to $\mathbb{R}^{n}$ and the same hold
for their dual $\mathbb{V}^{\ast}$ (see Appendix). The symbol
$\cdot $ indicates the scalar product of two objects having the
same structure: two tensors, two scalar variables or two\
collections of tensorial and/or scalar variables (the same
notation has been used e.g. by \citet[pg. 428] {JirasekBazant02}).
The inequality (\ref{2princ2}) can be written in a slightly
different manner, by introducing the \emph{dissipative}
thermodynamic forces
\begin{equation}
\mathbf{q}^{d}=\left[ \mbox{\boldmath
$\sigma$}^{d},\mbox{\boldmath $\tau$}_{1}^{d},..., \mbox{\boldmath
$\tau$}_{N}^{d}\right] :=\left[ \mbox{\boldmath
$\sigma$}-\mbox{\boldmath $\sigma$} ^{nd},-\mbox{\boldmath
$\tau$}_{1}^{nd},...,-\mbox{\boldmath $\tau$}_{N}^{nd}\right] \in
\mathbb{V}^{\ast }  \label{sigSigd}
\end{equation}
Hence,
\begin{equation}
\Phi _{m}\left( t\right) =\mbox{\boldmath
$\sigma$}^{d}:\mbox{\boldmath $\dot{\varepsilon}$}+
\sum_{i=1}^{N}\mbox{\boldmath $\tau$}_{i}^{d}\cdot \mbox{\boldmath
$\dot{\chi}$}_{i}\mathbf{=q} ^{d}\cdot \mathbf{\dot{v}}\geq 0
\label{2ndprinc1}
\end{equation}

The forces $\mathbf{q}^{d}$ have to be defined in such a way that
the couple $\left( \mathbf{q}^{d},\mathbf{\dot{v}}\right) $ always
fulfils the inequality (\ref{2ndprinc1}). Therefore, some
additional \emph{complementary rules} have to be introduced. They
can be defined by assuming the existence of a non-negative,
proper, convex and lower semi-continuous function $\phi
:\mathbb{V}\rightarrow (-\infty ,\infty ]$ (Appendix, item 2),
called \emph{pseudo-potential}, in general non-differentiable,
such that $\phi \left( \mathbf{0}\right) =0$ and:
\begin{equation}
\mathbf{q}^{d}\in \partial \phi \left( \mathbf{\dot{v}}\right)
\label{Normality}
\end{equation}
where $\partial $ indicates the sub-differential operator (see
Appendix, item 3). This condition is called \emph{generalized
normality}. A more detailed way of writing (\ref{Normality}) is
\begin{equation}
\mathbf{q}^{d}\in \left.\partial \phi \left(
\mathbf{\dot{v}^{\prime}};\mathbf{v}\right)\right\vert_{\mathbf{\dot{v}^{\prime}}=\mathbf{\dot{v}}}
\label{NormalityDetail}
\end{equation}
As a matter of fact, $ \phi $ is a general function of the fluxes
$\mathbf{\dot{v}^{\prime}}$ and may also depend on the state
variables $\mathbf{v}$. However, the subdifferential is taken, by
definition, only with respect to the fluxes $\mathbf{
\dot{v}^{\prime}}$ and the thermodynamic force $\mathbf{q}^{d}$
corresponds to the subdifferential of $\phi$ at $\mathbf{
\dot{v}^{\prime}}=\mathbf{ \dot{v}}$, where $\mathbf{ \dot{v}}$ is
the \emph{actual} flow. By using the properties of
sub-differentials, it can be proved that for dissipative forces
defined by (\ref {Normality}), the inequality $\mathbf{q}^{d}\cdot
\mathbf{\dot{v}\geq }$ $0$ is always fulfilled (Appendix, item 4).
Therefore, the second principle (\ref{2ndprinc1}) is also
satisfied.

A dual pseudo-potential $\phi ^{\ast }:\mathbb{V}^{\ast
}\rightarrow (-\infty ,\infty ]$ can be defined by the
Legendre-Fenchel transform of $ \phi $:
\begin{equation}
\phi ^{\ast }\left( \mathbf{q}^{d^{\prime}}\right)
:=\sup_{\mathbf{\dot{v}^{\prime}\in } \mathbb{V}}\left(
\mathbf{q}^{d^{\prime}} \cdot \mathbf{\dot{v}^{\prime}}-\phi
\left( \mathbf{\dot{v}^{\prime} }\right) \right)  \label{LegFench}
\end{equation}
When $\phi $ has an additional dependence on the state variables
$\mathbf{v}$ , then (\ref{LegFench}) leads to $\phi ^{\ast }=\phi
^{\ast }\left( \mathbf{q}^{d^{\prime}}\mathbf{;v}\right) $. It can
be proved that the dual pseudo-potential is a non-negative,
proper, convex and lower semi-continuous function of
$\mathbf{q}^{d^{\prime}}$, such that $\phi ^{\ast }\left(
\mathbf{0}\right) =0$ (see Appendix, item 5). The dual normality
condition reads
\begin{equation}
\mathbf{\dot{v}}\in \partial \phi ^{\ast }\left(
\mathbf{q}^{d}\right) \label{NormalityDual}
\end{equation}
where $\mathbf{q}^{d}$ is the actual value of the dissipative
force. The expression (\ref{NormalityDual}) is equivalent to
\begin{equation}
\mathbf{\dot{v}}\left. \in \partial \phi ^{\ast }\left(
\mathbf{q}^{d^{\prime}};\mathbf{v}\right)\right\vert_{\mathbf{q}^{d^{\prime}}=\mathbf{q}^{d}}
\label{NormalityDualDetail}
\end{equation}
and it guarantees that $\mathbf{q}^{d}\cdot\mathbf{\dot{v}}\geq 0$
(Appendix, item 5). Moreover, it defines the complementarity rules
of \emph{generalized standard materials} \citep{Halphen75},
sometimes called \emph{fully associated materials} \cite[pg. 452]
{JirasekBazant02}.

Plasticity is characterized by a \emph{rate-independent} memory
effect \citep[pg. 13]{Visintin94}. This special behavior occurs
when the pseudo-potential $\phi $ is a positively homogeneous
function of order 1 with respect to the fluxes
$\mathbf{\dot{v}}^{\prime}$. In this case, provided that
$\mathbf{q}^{d}$ is computed from (\ref{Normality}) or that
$\mathbf{\dot{v}}$ derives from (\ref{NormalityDual}), it can be
proved that the pseudo-potential at $\mathbf{\dot{v}}$ is equal to
the intrinsic dissipation, viz.
$\phi\left(\mathbf{\dot{v}}\right)=\mathbf{q}^{d}\cdot
\mathbf{\dot{v}= } \ \Phi _{m}$ (Appendix, item 6). Moreover, the
dual pseudo-potential $\phi ^{\ast }$ becomes the indicator
function of a closed convex set $\bar{\mathbb{E}} \subset
\mathbb{V}^{\ast }$ and the normality rule (\ref{NormalityDual})
entails that, given the dissipative force $\mathbf{q} ^{d}\in
\bar{\mathbb{E}}$, the flux $\mathbf{\dot{v}}$ fulfils the
following condition:
\begin{equation}
\begin{array}{c}
\forall \mathbf{q}^{d^{\prime}}\in \bar{\mathbb{E}} \ \ \ \ \
\left( \mathbf{q}^{d^{\prime}}-\mathbf{q}^{d}\right) \cdot
\mathbf{\dot{v}}\leq 0
\end{array}
\label{orthog}
\end{equation}
viz. for a given dissipative force $\mathbf{q}^{d}$, the flow
$\mathbf{\dot{v}}$ defined by (\ref{orthog}) (or, equivalently, by
(\ref{NormalityDual}) or (\ref{NormalityDualDetail})) is such that
its power when it is associated to the actual force
$\mathbf{q}^{d}$ is always greater or equal to the power
$\mathbf{q}^{d^{\prime}}\cdot \mathbf{\dot{v}}$ of all the other
dissipative forces $\mathbf{q}^{d^{\prime}}\in\bar{\mathbb{E}}$
(\emph{generalized maximum-dissipation principle}
\citep{Halphen75}). When $\mathbf{q}^{d}\in
\partial \bar{\mathbb{E}}$, the inequality (\ref{orthog}) indicates that
$\mathbf{\dot{v}}$ belongs to the cone orthogonal to $\partial
\bar{\mathbb{E}}$ at the point $\mathbf{q}^{d}$. When
$\mathbf{q}^{d}\in int\left( \bar{\mathbb{E}}\right) $, it forces
the flow $\mathbf{\dot{v}}$ to be zero (Appendix, item 7).

\section{Prandtl-Reuss model}

\subsection{Perfectly plastic Prandtl-Reuss model}

In order to illustrate the general procedure that is adopted
hereinafter, the basic example of the Prandtl-Reuss model is
considered first. The relevant state variables are the total and
the plastic strain $\mathbf{v} =\left(\mbox{\boldmath
$\varepsilon$},\mbox{\boldmath $\varepsilon$}^{p}\right) $ and
$\mathbf{q} ^{nd}=\left( \mbox{\boldmath
$\sigma$}^{nd},\mbox{\boldmath $\tau$}^{nd}\right) $ are the
associated non-dissipative thermodynamic forces. The usual
quadratic form of the Helmholtz free energy density $\Psi $ is
used, in order to preserve the linear dependence of all
non-dissipative forces with respect to state variables:
\begin{equation}
\Psi =\frac{1}{2}\left(\mbox{\boldmath $\varepsilon
-\varepsilon$}^{p}\right) : \mathbf{C}:\left( \mbox{\boldmath
$\varepsilon -\varepsilon$}^{p}\right) \label{PsiReuss}
\end{equation}
For isotropic materials, the fourth-order tensor of the elastic
moduli is equal to $\mathbf{C=}\left( K-\frac{2}{3}G\right)
\mathbf{1\otimes 1+}2G \mathbf{I}$, where $K$ is the (isothermal)
bulk modulus, $G$ is the shear modulus and $\mathbf{\otimes }$ is
the direct (or outer) product of two second order tensors. The
assumption of isotropy is always adopted, even if the concise
symbol $\mathbf{C}$ is used. The non-dissipative forces associated
with (\ref {PsiReuss}) can be derived by means of
(\ref{fnondiss}):
\begin{equation}
\begin{array}{c}
\mbox{\boldmath $\sigma$}^{nd}=\mathbf{C}:\left( \mbox{\boldmath
$\varepsilon-\varepsilon$} ^{p}\right) ,\ \ \ \ \ \ \ \ \
\mbox{\boldmath $\tau$}^{nd}=-\mathbf{C}:\left( \mbox{\boldmath
$\varepsilon-\varepsilon$}^{p}\right)
\end{array}
\label{sigSimpl}
\end{equation}
The evolution of the dissipative forces $\mathbf{q} ^{d}=\left(
\mbox{\boldmath $\sigma$}^{d},\mbox{\boldmath $\tau$}^{d}\right)
\in \mathbb{S} ^{2}\times \mathbb{S}^{2}:=\mathbb{V}^{\ast }$ is
defined by introducing a suitable pseudo-potential $\phi $, which
is a function of the fluxes $ \mathbf{\dot{v}}^{\prime}=\left(
\mbox{\boldmath $\dot{\varepsilon}$}^{\prime},\mbox{\boldmath
$\dot{\varepsilon}$}^{p^{\prime}}\right) \in \mathbb{S}^{2}\times
\mathbb{S}^{2}:=\mathbb{V}$ ($\times $ is the cartesian product):
\begin{equation}
\begin{array}{l}
\phi \left(\mbox{\boldmath
$\dot{\varepsilon}$}^{\prime},\mbox{\boldmath
$\dot{\varepsilon}$}^{p^{\prime}} \right)=\sqrt{\frac{2}{3}}\sigma
_{y}\ \left\Vert \mbox{\boldmath
$\dot{\varepsilon}$}^{p^{\prime}}\right\Vert
+\mathbb{I}_{\bar{\mathbb{D}}}\left( \mbox{\boldmath
$\dot{\varepsilon}$}^{\prime},\mbox{\boldmath
$\dot{\varepsilon}$}^{p^{\prime}}\right)  \\
\textrm{}\bar{\mathbb{D}}=\left\{ \left( \mbox{\boldmath
$\dot{\varepsilon}$}^{\prime},\mbox{\boldmath
$\dot{\varepsilon}$}^{p^{\prime}}\right) \mathbf{\in } \
\mathbb{V} \textrm{ \ \ \ \ such \ that \ \ }tr\left(
\mbox{\boldmath $\dot{\varepsilon}$}^{p^{\prime}}\right)
=0\right\}
\end{array}
\label{PseudoReuss}
\end{equation}
where $tr(\mathbf{u})$ indicates the trace of the tensor
$\mathbf{u\in }\mathbb{S}^{2}$. The norm of the second order
symmetric tensor $\mathbf{u}$ is given by $\left\Vert \mathbf{u}
\right\Vert =\sqrt{u_{ij} \ u_{ij}}$. If in addition $tr\left(
\mathbf{u}\right) =u_{ii}=0$, then $\left\Vert
\mathbf{u}\right\Vert^{2} =2J_{2}\left( \mathbf{u}\right) $ where
$J_{2}\left( \mathbf{u}\right) $ is the second invariant of the
deviatoric part of $\mathbf{u}$; $\sigma _{y}$ is the
one-dimensional tension stress limit and
$\mathbb{I}_{\bar{\mathbb{D}}}$ is the indicator function of the
set $\bar{\mathbb{D}}$, namely $\mathbb{I}_{\bar{\mathbb{D}}}=0$
if $tr\left( \mbox{\boldmath $\dot{\varepsilon}$}
^{p^{\prime}}\right) =0$ and
$\mathbb{I}_{\bar{\mathbb{D}}}=+\infty $ elsewhere. This set is
the effective domain of $\phi $ (Appendix, item 2). The
pseudo-potential $\phi $ is a homogeneous function of order 1 with
respect to $ \left(\mbox{\boldmath
$\dot{\varepsilon}$}^{\prime},\mbox{\boldmath
$\dot{\varepsilon}$}^{p^{\prime}}\right) $ and therefore a rate
independent constitutive behavior is expected and the dissipation
$\Phi_{m}$ is equal to
$\sqrt{\frac{2}{3}}\sigma_{y}\|\mbox{\boldmath
$\dot{\varepsilon}$}^{p}\|$, where $\mbox{\boldmath
$\dot{\varepsilon}$}^{p}$ is the actual plastic flow (Appendix,
item 6). The indicator function $ \mathbb{I}_{\bar{\mathbb{D}}}$
accounts for the fact that plastic deformation occurs without
volume changes (\emph{plastic incompressibility}). This assumption
is usual for metals and has been validated by experimental
evidence.

The pseudo-potential $\phi ^{\ast }$, dual of $\phi $, can be
computed using the Legendre-Fenchel transform (Appendix, item 5)
and is equal to:
\begin{equation}
\begin{array}{l}
\phi ^{\ast }\left(\mbox{\boldmath
$\sigma$}^{d^{\prime}},\mbox{\boldmath
$\tau$}^{d^{\prime}}\right)=\sup_{\left(\mbox{\scriptsize\boldmath
$\dot{\varepsilon}$}^{\prime},\mbox{\scriptsize\boldmath
$\dot{\varepsilon}$}^{p^{\prime}}\right) \mathbf{\in
}\bar{\mathbb{D}}}\left( \mbox{\boldmath $\sigma$}^{d^{\prime}}:
\mbox{\boldmath $\dot{\varepsilon}$}^{\prime}+\mbox{\boldmath
$\tau$}^{d^{\prime}}:\mbox{\boldmath
$\dot{\varepsilon}$}^{p^{\prime}}-\phi \right) \\
\ \ \ \ \ \ \  \ \ \ \ \ \ \ \ \ \
=\mathbb{I}_{\bar{\mathbb{E}}}\left( \mbox{\boldmath
$\sigma$}^{d^{\prime}},\mbox{\boldmath $\tau$}
^{d^{\prime}}\right)
\end{array}
\label{fistarReuss1}
\end{equation}
The dual pseudo-potential $\phi ^{\ast }$ is the indicator
function of a closed convex set $\bar{\mathbb{E}}$. Hence,
$\phi=\phi^{\ast\ast}$ is the support function of the same set
(Appendix, item 6). Moreover, since $\phi$ does not depend on
$\mbox{\boldmath $\dot{\varepsilon}$}^{\prime}$, the dual
pseudo-potential $\phi^{\ast}$ can be written as the sum of two
indicator functions (Appendix, item 7):
\begin{equation}
\begin{array}{l}
\phi^{\ast}\left(\mbox{\boldmath
$\sigma$}^{d^{\prime}},\mbox{\boldmath $\tau$}^{d^{\prime}}\right)
=\mathbb{I}_{0}\left( \mbox{\boldmath
$\sigma$}^{d^{\prime}}\right) +\mathbb{I}_{ \mathbb{E}}\left(
\mbox{\boldmath $\tau$}^{d^{\prime}}\right) \\
\mathbb{E}=\left\{ \mbox{\boldmath $\tau$}^{d^{\prime}}\in
\mathbb{S}^{2}\textrm{\ such \ that \ }f\left( \mbox{\boldmath
$\tau$}^{d^{\prime}}\right)=\left\Vert dev\left( \mbox{\boldmath
$\tau$}^{d^{\prime}}\right) \right\Vert - \sqrt{\frac{2}{3}}\sigma
_{y}\leq 0\right\}
\end{array}
\label{fistarReuss2}
\end{equation}
where $dev(\mathbf{u})$ is the deviatoric part of
$\mathbf{u\in}\mathbb{S}^{2}$. The first term is equivalent to the
condition $\mbox{\boldmath $\sigma$}^{d^{\prime}}=\mathbf{0}$,
while the other indicator function $\mathbb{I}_{\mathbb{E}}$
defines a region in the $\mbox{\boldmath $\tau$} ^{d^{\prime}}$
stress space. Recalling that the \emph{actual value} of
$\mbox{\boldmath $\tau$}^{d^{\prime}}$, viz. $\mbox{\boldmath
$\tau$}^{d}$, fulfils the condition $\mbox{\boldmath
$\tau$}^{d}=-\mbox{\boldmath $\tau$}^{nd}$ and that the only
possible value for $\mbox{\boldmath $\sigma$}^{d}$ is zero, using
(\ref{sigSimpl}) it is straightforward to see that $
\mbox{\boldmath $\tau$}^{d}=\mbox{\boldmath$
\sigma$}=\mbox{\boldmath $\sigma$}^{nd}$ and that $\mathbb{E}$ can
also be interpreted as a set in the $\mbox{\boldmath $\sigma$}$
stress space. The associated function $f$ is known as
\emph{loading function} and the condition $f=0$ defines the
\emph{plastic states}. The interior of $\mathbb{E}$ is associated
with the \emph{elastic states} and the whole (closed) set
$\mathbb{E}$ contains all \emph{plastically admissible states}.
The actual flows $\left( \mbox{\boldmath
$\dot{\varepsilon}$},\mbox{\boldmath
$\dot{\varepsilon}$}^{p}\right)$ can now be derived from
(\ref{fistarReuss2}) by computing the subdifferential set of $\phi
^{\ast }$ and then considering it at $\left( \mbox{\boldmath
$\sigma$}^{d^{\prime}},\mbox{\boldmath
$\tau$}^{d^{\prime}}\right)=\left( \mbox{\boldmath
$\sigma$}^{d},\mbox{\boldmath $\tau$}^{d}\right)$. Hence, no
restrictive conditions are imposed on $\mbox{\boldmath
$\dot{\varepsilon}$}$, while the plastic strain flow reads
(Appendix, item 7)
\begin{equation}
\begin{array}{l}
\mbox{\boldmath $\dot{\varepsilon}$}^{p}=\frac{dev\left(
\mbox{\boldmath $\tau$}^{d}\right) }{\left\Vert dev\left(
\mbox{\boldmath $\tau$}^{d}\right) \right\Vert }\dot{\lambda}=
\mathbf{n}\ \dot{\lambda}\textrm{ \ \ \ \ \ \ \ with \ } \\
\dot{\lambda}\geq 0, \ \ \ f\left( \mbox{\boldmath
$\tau$}^{d}\right)\leq 0, \ \ \ \ \dot{\lambda}f\left(
\mbox{\boldmath $\tau$}^{d}\right)=0
\end{array}
\label{FlowReuss}
\end{equation}
Observe that $f$ in the loading-unloading conditions in the second
row of (\ref{FlowReuss}) is computed at the actual stress state.
The plastic multiplier $\dot{\lambda}$ is then evaluated by
imposing the \emph{consistency condition}, i.e.
$\dot{\lambda}\dot{f} =0$ (see e.g. \cite{Simo88}), with
\begin{equation}
\dot{f}=\left[ \frac{\partial f}{\partial \mbox{\boldmath
$\tau$}^{d^{\prime}}}:\mbox{\boldmath
$\dot{\tau}$}^{d^{\prime}}\right]_{\mbox{\boldmath
$\tau$}^{d^{\prime}}=\mbox{\boldmath $\tau$}^{d}}
\end{equation}
Note that $\dot{f}$ is computed from the general expression of
$f\left(\mbox{\boldmath $\tau$}^{d^{\prime}}\right)$ and then
evaluated at the actual state $\mbox{\boldmath
$\tau$}^{d^{\prime}}=\mbox{\boldmath $\tau$}^{d}$. Consistency
corresponds to the requirement that in order to have
$\dot{\lambda}>0$, the \emph{actual} dissipative force
$\mbox{\boldmath $\tau$}^{d}\in\partial \mathbb{E}$ cannot leave
$\partial \mathbb{E}$ during the plastic flow. Hence, by using
(\ref{FlowReuss}) and the relationship $\mbox{\boldmath
$\tau$}^{d}=\mbox{\boldmath
$\sigma$}=\mathbf{C}:\left(\mbox{\boldmath
$\varepsilon$}-\mbox{\boldmath $\varepsilon$}^{p}\right)$, one has
\begin{equation}
\dot{f}=\frac{dev\left( \mbox{\boldmath $\tau$}^{d}\right)
}{\left\Vert dev\left( \mbox{\boldmath $\tau$}^{d}\right)
\right\Vert }:dev\left(\mbox{\boldmath $\dot{\tau}$}^{d}\right)
=\mathbf{n}:\mathbf{C}:\mbox{\boldmath
$\dot{\varepsilon}$}\mathbf{-n}:\mathbf{C}: \mathbf{n} \
\dot{\lambda}
\end{equation}
thus $\dot{\lambda}=H\left( f\right) \frac{\left\langle \mathbf{
n:C}:\mbox{\footnotesize\boldmath
$\dot{\varepsilon}$}\right\rangle }{\mathbf{n:C:n}}=H\left(
f\right) \left\langle \mathbf{n}:\mbox{\boldmath
$\dot{\varepsilon}$}\right\rangle $ , where $ \left\langle
x\right\rangle =\frac{x+\left\vert x\right\vert }{2}$
(\emph{McCauley brackets}) and $H\left( f\right) $ is the
Heaviside function, equal to zero for $f<0$ and equal to 1
elsewhere.

In summary, the Prandtl-Reuss perfectly plastic model was
formulated by means of the Helmholtz free energy $\Psi $ and the
pseudo-potential $\phi $; then, the dual potential $\phi ^{\ast }$
was computed from the Legendre-Fenchel transform of $\phi $; the
subdifferential set of $\phi ^{\ast }$ was used to define the
fluxes and the consistency assumption led to the determination of
the plastic multiplier $\dot{\lambda}$. In the next sections, this
approach will be used to formulate two Prandtl-Reuss models with
isotropic hardening and other more complex plasticity models, such
as endochronic, NLK hardening and generalized plasticity models.
In order to get this result, some non-standard expressions for the
pseudo-potentials $ \phi $ and $\phi ^{\ast }$ are introduced.

\subsection{Classical Prandtl-Reuss model with isotropic hardening\label{SecPraReuss}}

The vector of the representative state variables for a
Prandtl-Reuss model with isotropic hardening is equal to
$\mathbf{v}=\left(\mbox{\boldmath $\varepsilon$},\mbox{\boldmath
$\varepsilon$}^{p},\zeta \right)$, while
$\mathbf{q}^{nd}=\left(\mbox{\boldmath
$\sigma$}^{nd},\mbox{\boldmath $\sigma$},R^{nd}\right) $  are the
associated non-dissipative forces. The scalar internal variable
$\zeta $ and the associated forces $R^{d}$ and $R^{nd}$ are
introduced in order to represent the isotropic hardening.
Moreover, $\mathbf{\dot{v}}=\left(\mbox{\boldmath
$\dot{\varepsilon}$},\mbox{\boldmath
$\dot{\varepsilon}$}^{p},\dot{\zeta}\right) \in
\mathbb{V=S}^{2}\times \mathbb{ S}^{2}\times \mathbb{R}$ is the
flux vector and $\mathbf{q}^{d}=\left(\mbox{\boldmath
$\sigma$}^{d},\mbox{\boldmath $\sigma$}^{d},R^{d}\right) \in
\mathbb{V}^{\ast } \mathbb{=S}^{2}\times \mathbb{S}^{2}\times
\mathbb{R}$ contains all dissipative thermodynamic forces. The
Helmholtz free energy is assumed to be of the form
\begin{equation}
\Psi =\frac{1}{2}\left(\mbox{\boldmath $\varepsilon -\varepsilon$
}^{p}\right) : \mathbf{C}:\left(\mbox{\boldmath $\varepsilon
-\varepsilon$}^{p}\right) +\xi \left( \zeta \right)
\label{PsiClassIso}
\end{equation}
where $\xi \left( \zeta \right) $ is a scalar function such that
$\xi \left( 0\right) =0$ and $\frac{d\xi}{d\zeta} \left( 0\right)
=0$. It follows that
\begin{equation}
\begin{array}{c}
\mbox{\boldmath $\sigma$}^{nd}=\mathbf{C}:\left( \mbox{\boldmath
$\varepsilon -\varepsilon$} ^{p}\right) , \ \ \ \ \ \ \
\mbox{\boldmath $\tau$}^{nd}=-\mathbf{C}:\left( \mbox{\boldmath
$\varepsilon -\varepsilon$}^{p}\right) , \ \ \ \ \ \ \
R^{nd}=\frac{d\xi}{d\zeta}\left( \zeta \right)
\end{array}
\label{J2isoNonDissip}
\end{equation}
The pseudo-potential is assumed of the following form:
\begin{equation}
\begin{array}{l}
\phi\left( \mbox{\boldmath
$\dot{\varepsilon}$}^{\prime},\mbox{\boldmath
$\dot{\varepsilon}$}^{p^{\prime}},\dot{\zeta}^{\prime}\right)
=\sqrt{\frac{2}{3}}\sigma _{y} \
\dot{\zeta}^{\prime}+\mathbb{I}_{\bar{\mathbb{D}} }\left(
\mbox{\boldmath $\dot{\varepsilon}$}^{\prime},\mbox{\boldmath
$\dot{\varepsilon}$}^{p^{\prime}},\dot{\zeta}^{\prime}\right)
\\
\\
\bar{\mathbb{D}}\mathbf{=}\left\{
\begin{array}{l}
\left( \mbox{\boldmath
$\dot{\varepsilon}$}^{\prime},\mbox{\boldmath
$\dot{\varepsilon}$}^{p^{\prime}},\dot{\zeta}^{\prime}\right) \in
\mathbb{V}\textrm{ \ such \ that}  \ \ \ \ tr\left(
\mbox{\boldmath $\dot{\varepsilon}$}^{p^{\prime}}\right)
=0\textrm{ \ \ and \ \ }\dot{\zeta^{\prime}}\geq \left\Vert
\mbox{\boldmath $\dot{\varepsilon}$}^{p^{\prime}}\right\Vert
\end{array}
\right\}
\end{array}
\label{PseudoReussIso}
\end{equation}
The first term in the expression of $\phi$ is the same as in the
perfectly-plastic model when $\dot{\zeta}^{\prime}$ is equal to
the norm of $\mbox{\boldmath $\dot{\varepsilon}$}^{p^{\prime}}$.
The second term, that is the indicator function $
\mathbb{I}_{\bar{\mathbb{D}}}$, depends not only on
$tr(\mbox{\boldmath $\dot{\varepsilon}$}^{p^{\prime}})$, but also
on the flow $\dot{\zeta^{\prime} }$, which is forced be greater or
equal than the norm of the plastic strain flow. This inequality
guarantees that $\dot{\zeta^{\prime}}$ and $\phi$ are non-negative
and entails that $\bar{\mathbb{D}}$ is convex and closed (see
Appendix, item 1 and Figure \ref{FigPrandtl1}a, which illustrates
the projection of $\bar{\mathbb{D}}$ on the $\left(\mbox{\boldmath
$\dot{\varepsilon}$}^{p^{\prime}},\dot{\zeta}^{\prime}
\right)$-plane for the tension-compression case). The dual
pseudo-potential is different from (\ref{fistarReuss1}), due to
the presence of the dissipative force $R^{d^{\prime}}$ associated
with $\dot{\zeta}^{\prime}$:
\begin{equation}
\begin{array}{l}
\phi ^{\ast }\left(\mbox{\boldmath
$\sigma$}^{d^{\prime}},\mbox{\boldmath
$\tau$}^{d^{\prime}},R^{d^{\prime}}\right)=\sup_{\left(
\mbox{\scriptsize\boldmath
$\dot{\varepsilon}$}^{\prime},\mbox{\scriptsize\boldmath
$\dot{\varepsilon}$}^{p^{\prime}}, \dot{\zeta}^{\prime}\right)
\mathbf{\in }\bar{\mathbb{D}}}\left(\mbox{\boldmath
$\sigma$}^{d^{\prime}}: \mbox{\boldmath
$\dot{\varepsilon}$}^{\prime}+\mbox{\boldmath
$\tau$}^{d^{\prime}}:\mbox{\boldmath
$\dot{\varepsilon}$}^{p^{\prime}}+R^{d^{\prime}}\dot{\zeta}^{\prime}-\phi \right) \\
=\mathbb{I}_{0}\left(\mbox{\boldmath $\sigma$}^{d^{\prime}}\right)
+\mathbb{I}_{\mathbb{E}}\left(\mbox{\boldmath
$\tau$}^{d^{\prime}},R^{d^{\prime}}\right)
\end{array}\label{ElDom_iso}
\end{equation}
where $ \mathbb{E}=\left\{ \left(\mbox{\boldmath
$\tau$}^{d^{\prime}},R^{d^{\prime}}\right) \in
\mathbb{S}^{2}\mathbb{\times R} \textrm{\ \ such \ that} \ \ \
f\left(\mbox{\boldmath
$\tau$}^{d^{\prime}},R^{d^{\prime}}\right)\leq 0 \right\}$  and
\begin{equation}
f\left(\mbox{\boldmath
$\tau$}^{d^{\prime}},R^{d^{\prime}}\right)=\left\Vert
dev\left(\mbox{\boldmath $\tau$}^{d^{\prime}}\right) \right\Vert
-\left( \sqrt{\frac{2}{3}}\sigma _{y}-R^{d^{\prime}}\right) \leq 0
\label{ElDom_iso1}
\end{equation}
The loading function $f$ defines a convex and closed region
$\mathbb{E}$ in the $\left( \mbox{\boldmath
$\tau$}^{d^{\prime}},R^{d^{\prime}}\right) $ space, where the
\emph{actual value} of $R^{d^{\prime}}$, viz. $
R^{d}=-R^{nd}=-\frac{d\xi\left( \zeta \right)}{d\zeta}$  governs
isotropic hardening (or softening). The limit stress becomes
greater than its initial value $\sqrt{\frac{2}{3}}\sigma_{y}$ when
$\frac{d\xi\left( \zeta \right)}{d\zeta} \geq 0$ and less when
$\frac{d\xi\left( \zeta \right)}{d\zeta} \leq 0$. Figure
\ref{FigPrandtl1}b illustrates the set $\mathbb{E}$. The flow
rules follow from the generalized normality conditions:
\begin{equation}
\begin{array}{l}
\mbox{\boldmath $\dot{\varepsilon}$}^{p}=\frac{dev\left(
\mbox{\footnotesize\boldmath $\tau$}^{d}\right) }{\left\Vert
dev\left(\mbox{\footnotesize\boldmath $\tau$}^{d}\right)
\right\Vert } \dot{\lambda}=
\mathbf{n}\dot{\lambda}, \ \ \ \ \ \ \dot{\zeta}=\dot{\lambda} \\
\textrm{with \ \ \ }\dot{\lambda}\geq 0, \ \ \ f\leq 0, \ \ \ \
\dot{\lambda}f=0
\end{array}
\label{flowReuss0}
\end{equation}
The flow of the internal variable $\zeta $ is equal to the plastic
multiplier $\dot{\lambda}$, which can be evaluated by imposing the
consistency condition:
\begin{equation}
\dot{f}=\left[ \frac{\partial f}{\partial \mbox{\boldmath
$\tau$}^{d^{\prime}}}:\mbox{\boldmath
$\dot{\tau}$}^{d^{\prime}}+\frac{\partial f}{\partial
R^{d^{\prime}}} \
\dot{R}^{d^{\prime}}\right]_{\left(\mbox{\boldmath
$\tau$}^{d^{\prime}}=\mbox{\boldmath
$\tau$}^{d},R^{d^{\prime}}=R^{d}\right)}=0
\end{equation}
It follows that
\begin{equation}
\dot{\lambda}=H\left( f\right) \frac{\left\langle
\mathbf{n:C}:\mbox{\boldmath $\dot{\varepsilon}$}\right\rangle
}{\mathbf{n:C:n+}\frac{d^{2}\xi\left( \zeta \right)
}{d\zeta^{2}}}=H\left( f\right) \frac{\left\langle
\mathbf{n}:\mbox{\boldmath $\dot{\varepsilon}$}\right\rangle
}{1\mathbf{+}\frac{1}{2G}\frac{d^{2}\xi\left( \zeta
\right)}{d\zeta^{2}} } \label{lambdaReuss}
\end{equation}
where $H\left( f\right)$ and $\langle \ \rangle$ still indicate
the Heaviside function and McCauley brackets, respectively.

\subsection{Modified Prandtl-Reuss model with isotropic hardening}

\begin{figure}
\begin{center}
\includegraphics[width=13.4cm]{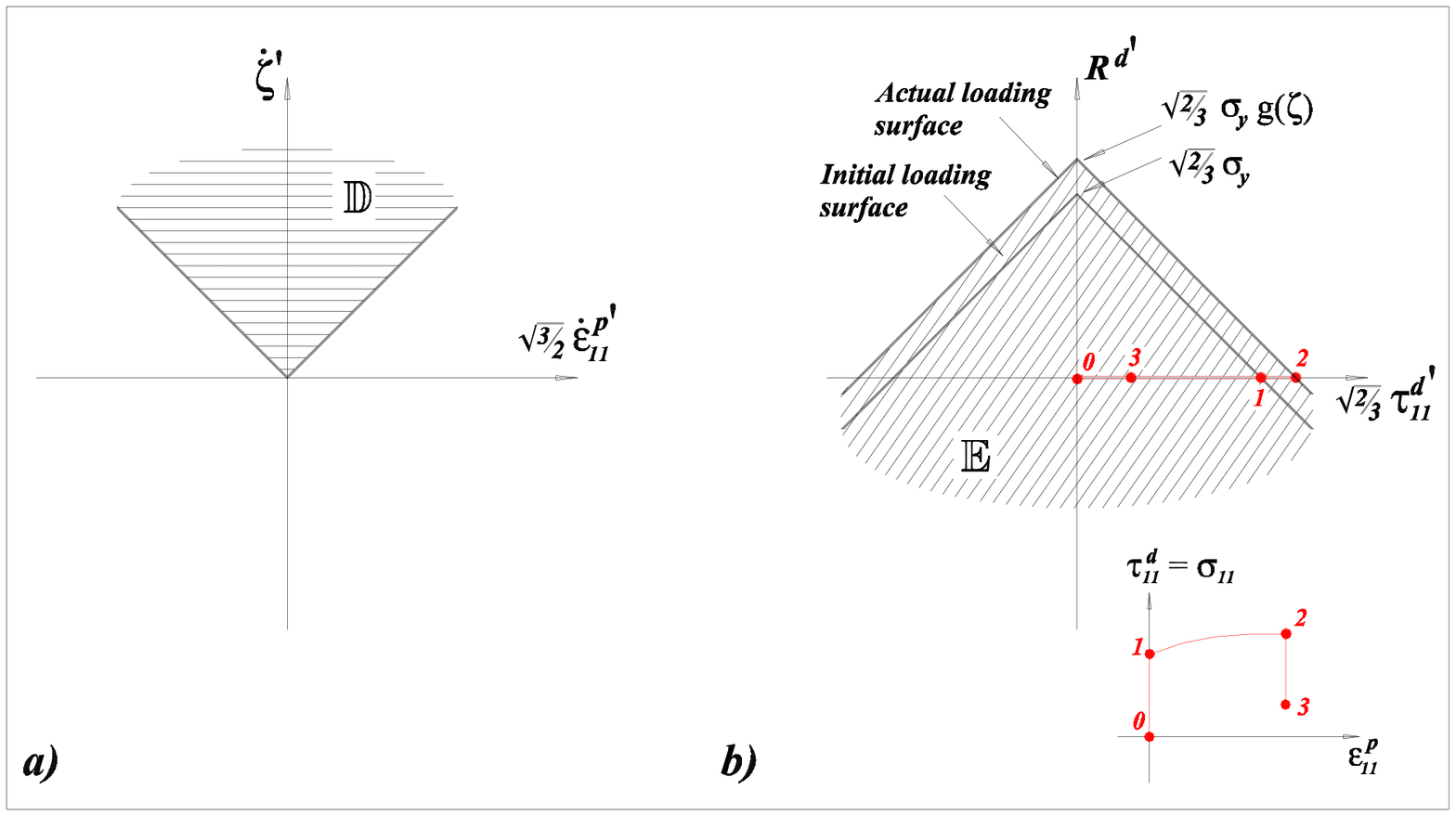}
\caption{Modified Prandtl-Reuss model. Tension-compression case
with $\xi(\zeta)=0$. \newline a) Projection of the
pseudo-potential effective domain $\bar{\mathbb{D}}$ on the
$\left(\mbox{\boldmath
$\dot{\varepsilon}$}^{p^{\prime}},\dot{\zeta}^{\prime}\right)$-plane.
This set is indicated by $\mathbb{D}$. b) Different configurations
of the domain $\mathbb{E}$. The position of $\mathbb{E}$ changes
according to the value of the internal variable $\zeta$. The point
$(\mbox{\boldmath $\tau$}^{d},R^{d})$, representing the actual
state, always lies on the axis $R^{d^{\prime}}=0$.}
\label{FigPrandtl2}
\end{center}
\end{figure}

The classical model of the previous section can be extended as
follows. Assume the state variables
$\mathbf{v}=\left(\mbox{\boldmath$\varepsilon$},\mbox{\boldmath
$\varepsilon$}^{p},\zeta \right)$ and let the Helmholtz energy be
equal to
\begin{equation}
\Psi =\frac{1}{2}\left(\mbox{\boldmath $\varepsilon -\varepsilon$
}^{p}\right) : \mathbf{C}:\left(\mbox{\boldmath $\varepsilon
-\varepsilon$}^{p}\right) +\xi \left( \zeta \right)
\label{PsiClassIsoTwice}
\end{equation}
As a result, the non-dissipative forces are the same as in Eq.
(\ref{J2isoNonDissip}). Then, a generalized definition of the
pseudo-potential $\phi $ is adopted:
\begin{equation}
\begin{array}{l}
\phi\left( \mbox{\boldmath
$\dot{\varepsilon}$}^{\prime},\mbox{\boldmath$\dot{\varepsilon}$}^{p^{\prime}}
,\dot{\zeta}^{\prime};\zeta \right) =\left(
\sqrt{\frac{2}{3}}\sigma _{y} g\left( \zeta \right)
-\frac{d\xi\left( \zeta \right)}{d\zeta} \right)
\dot{\zeta}^{\prime}+\mathbb{I}_{\bar{\mathbb{D}} }\left(
\mbox{\boldmath $\dot{\varepsilon}$}^{\prime},
\mbox{\boldmath$\dot{\varepsilon}$}^{p^{\prime}},\dot{\zeta}^{\prime}\right)
\\
\bar{\mathbb{D}}\mathbf{=}\left\{
\begin{array}{l}
\left( \mbox{\boldmath
$\dot{\varepsilon}$}^{\prime},\mbox{\boldmath$\dot{\varepsilon}$}^{p^{\prime}},\dot{\zeta}^{\prime}\right)
\in \mathbb{V}\textrm{ \ such \ that}
 \ \ tr\left( \mbox{\boldmath $\dot{\varepsilon}$}
^{p^{\prime}}\right) =0\textrm{ \ \ and \ \
}\dot{\zeta}^{\prime}\geq \left\Vert \mbox{\boldmath
$\dot{\varepsilon}$}^{p^{\prime}}\right\Vert
\end{array}
\right\}
\end{array}
\label{PseudoReussIso1}
\end{equation}
In this case, $\phi $ explicitly depends on the internal variable
$\zeta $, by means of $\frac{d\xi\left( \zeta \right)}{d\zeta} $\
and of the function $g\left( \zeta \right) $, positive and such
that $g\left( 0 \right)=1 $.  In the particular case where
$g\left( \zeta \right) =1+\frac{d\xi\left( \zeta \right)
}{d\zeta}\sqrt{\frac{3}{2}}\frac{1}{\sigma _{y}}$, the classical
expression given in Eq. (\ref{PseudoReussIso}) is recovered. The
dual pseudo-potential $\phi ^{\ast }$ can be evaluated from the
standard procedure, thus yielding:
\begin{equation}
\phi ^{\ast }\left( \mbox{\boldmath
$\sigma$}^{d^{\prime}},\mbox{\boldmath
$\tau$}^{d^{\prime}},R^{d^{\prime}};\zeta\right)=\mathbb{I}_{0}\left(
\mbox{\boldmath $\sigma$}^{d^{\prime}}\right)+
\mathbb{I}_{\mathbb{E}}\left( \mbox{\boldmath
$\tau$}^{d^{\prime}},R^{d^{\prime}};\zeta \right)
\label{fistarReussIso}
\end{equation}
where $\mathbb{E}=\left\{ \left( \mbox{\boldmath
$\tau$}^{d^{\prime}},R^{d^{\prime}}\right) \in \mathbb{S}^{2}
\mathbb{\times R}\textrm{\ \ \ \ such \ that \ \ \
}f\left(\mbox{\boldmath
$\tau$}^{d^{\prime}},R^{d^{\prime}};\zeta\right)\leq 0\right\}$
and
\begin{equation}
f\left(\mbox{\boldmath
$\tau$}^{d^{\prime}},R^{d^{\prime}};\zeta\right)=\left\Vert
dev\left( \mbox{\boldmath $\tau$} ^{d^{\prime}}\right) \right\Vert
-\left( \sqrt{\frac{2}{3}}\sigma _{y}g\left( \zeta \right)
-\frac{d\xi\left( \zeta \right)}{d\zeta} -R^{d^{\prime}}\right)
\label{fistarClassIso}
\end{equation}
In Figure \ref{FigPrandtl2}, the projection of $\bar{\mathbb{D}}$
on the $\left(\mbox{\boldmath
$\dot{\varepsilon}$}^{p^{\prime}},\dot{\zeta}^{\prime}
\right)$-plane and the set $\mathbb{E}$ are depicted for the
tension-compression case, with the assumption $\xi \left( \zeta
\right)=0$. The flow rules are the same as in the previous case
and they are reported below for completeness:
\begin{equation}
\begin{array}{l}
\mbox{\boldmath $\dot{\varepsilon}$}^{p}=\frac{dev\left(
\mbox{\footnotesize\boldmath $\tau$}^{d}\right) }{\left\Vert
dev\left(\mbox{\footnotesize\boldmath $\tau$}^{d}\right)
\right\Vert } \dot{\lambda}=
\mathbf{n}\dot{\lambda}, \ \ \ \ \ \ \dot{\zeta}=\dot{\lambda} \\
\textrm{with \ \ \ }\dot{\lambda}\geq 0, \ \ \ f\leq 0, \ \ \ \
\dot{\lambda}f=0
\end{array}
\label{flowReuss}
\end{equation}
In this case, $\dot{f}$ has to be computed accounting for the
state variables. Hence consistency condition reads
\begin{equation}
\dot{f}=\left[ \frac{\partial f}{\partial \mbox{\boldmath
$\tau$}^{d^{\prime}}}:\mbox{\boldmath
$\dot{\tau}$}^{d^{\prime}}+\frac{\partial f}{\partial
R^{d^{\prime}}} \ \dot{R}^{d^{\prime}}+ \frac{\partial f}{\partial
\zeta} \ \dot{\zeta}\right]_{\left(\mbox{\boldmath
$\tau$}^{d^{\prime}}=\mbox{\boldmath
$\tau$}^{d},R^{d^{\prime}}=R^{d}\right)}=0 \label{fdotzeta}
\end{equation}
and the plastic multiplier becomes equal to:
\begin{equation}
\dot{\lambda}=H\left( f\right) \frac{\left\langle
\mathbf{n:C}:\mbox{\boldmath $\dot{\varepsilon}$}\right\rangle
}{\mathbf{n:C:n+}\sqrt{\frac{2}{3}}\sigma _{y}\frac{dg\left( \zeta
\right) }{d\zeta}}=H\left( f\right) \frac{\left\langle
\mathbf{n}:\mbox{\boldmath $\dot{\varepsilon}$}\right\rangle
}{1\mathbf{+}\sqrt{\frac{2}{3}} \frac{\sigma
_{y}}{2G}\frac{dg\left( \zeta \right) }{d\zeta}}
\label{lambdaReussMod}
\end{equation}
provided that $1+\sqrt{\frac{2}{3}}\frac{\sigma
_{y}}{2G}\frac{dg\left( \zeta \right) }{d\zeta} >0$. This
condition does not prevent softening, which occurs when
$\frac{dg\left( \zeta \right) }{d\zeta} \leq 0$.

The comparison of Eqs. (\ref{ElDom_iso1}) and
(\ref{fistarClassIso}) proves to be very interesting. First, the
usual loading function only depends on the dissipative forces,
while $f$ in Eq. (\ref{fistarClassIso}) is also related to the
internal variable $\zeta $. Moreover, since
$R^{d}=-R^{nd}=-\frac{d\xi\left( \zeta \right)}{d\zeta} $, the
loading function (\ref{fistarClassIso}) at $\left(\mbox{\boldmath
$\tau$}^{d},R^{d}\right) $ becomes
\begin{equation}
f\left(\mbox{\boldmath $\tau$}^{d},R^{d};\zeta\right)=\left\Vert
dev\left( \mbox{\boldmath $\tau$} ^{d}\right) \right\Vert -
\sqrt{\frac{2}{3}}\sigma _{y}g\left( \zeta \right)\label{factual}
\end{equation}
This expression shows that the actual limit stress is equal to
$\sqrt{\frac{2}{3}}\sigma_{y}g\left( \zeta \right)$ and is
independent from the function $\xi \left( \zeta \right) $
introduced in the Helmholtz energy density (this is not the case
for the classical Prandtl-Reuss model).

\begin{figure}
\begin{center}
\includegraphics[width=13.4cm]{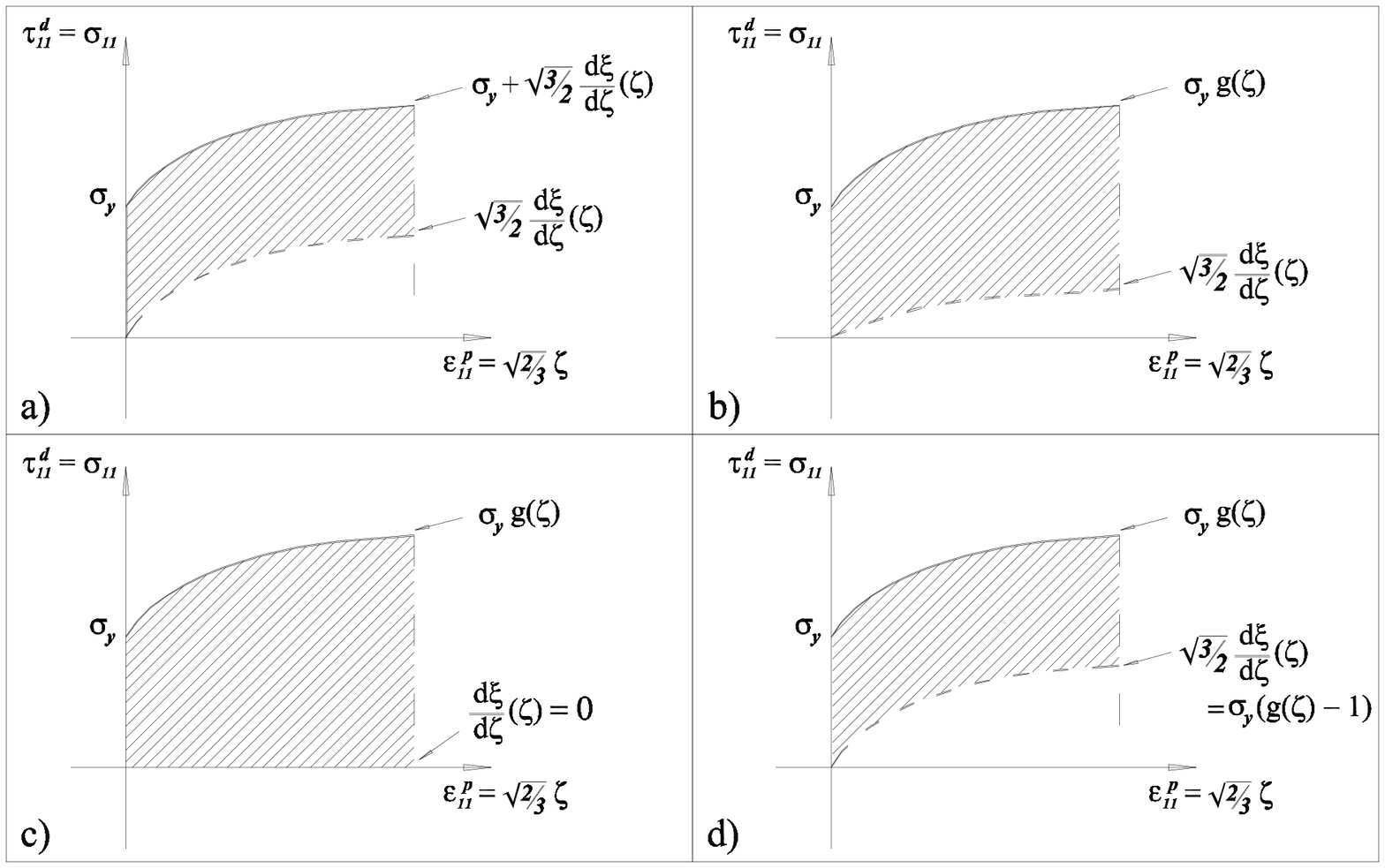}
\caption{Mechanical dissipation for the case of simple tension.
The hatched area is the energy $\int\Phi_{m}(t)dt$ dissipated
during the monotonic loading. \newline a) Classical Prandtl-Reuss
model. b) Modified Prandtl-Reuss model. c) Modified Prandtl-Reuss
model with $\xi(\zeta)=\frac{d\xi}{d\zeta}(\zeta)=0$. d) Modified
Prandtl-Reuss model with
$\sqrt{\frac{3}{2}}\frac{d\xi}{d\zeta}(\zeta)=\sigma_{y}(g(\zeta)-1)$:
the classical model is recovered.}

\label{FigDissPrandtl}
\end{center}
\end{figure}

The difference between the two Prandtl-Reuss models can be also
explained in terms of mechanical dissipation $\Phi_{m}$. For the
modified Prandtl-Reuss model, it is equal to
\begin{equation}
\Phi _{m}=\left( \sqrt{\frac{2}{3}}\sigma _{y}g\left( \zeta
\right) -\frac{d\xi\left( \zeta \right)}{d\zeta} \right)
\dot{\zeta}
\end{equation}
which is non-negative provided that $\frac{d\xi\left( \zeta
\right)}{d\zeta} \leq \sqrt{\frac{2}{3}}\sigma _{y} \ g\left(
\zeta \right) $. The case of a mono-dimensional monotonic loading
is depicted in Figure \ref{FigDissPrandtl}. The standard
Prandtl-Reuss model is characterized by the fact that the energy
$R^{d} \dot{\zeta}$ associated to isotropic hardening is not
dissipated. For this reason it is sometimes referred as
\emph{energy blocked in dislocations} \citep[pg.
402]{Lemaitre90engl}. Hence, the mechanical dissipation is equal
to $\sqrt{\frac{2}{3}}\sigma_{y} \dot{\zeta}$ for any function
$\xi(\zeta)$. Conversely, for the modified Prandtl-Reuss model the
amount of mechanical dissipation depends, for a given function
$g(\zeta)$, on the choice of $\xi \left( \zeta \right) $. Figure
\ref{FigDissPrandtl}b reports the case of generic functions
$g(\zeta)$ and $\xi(\zeta)$. Figure \ref{FigDissPrandtl}c and
\ref{FigDissPrandtl}d correspond to $\xi(\zeta)=0$ and to the case
where the modified model is equal to the classical one,
respectively.

\subsection{Multi-layer models of Prandtl-Reuss type\label{SecmultiPrandtl}}

Modified Prandtl-Reuss models, defined by Eqs.
(\ref{PsiClassIsoTwice})-(\ref{PseudoReussIso1}), can be directly
extended to multi-layer models \citep{Besseling58}. They consist
of a system of $N$ elastoplastic elements connected in parallel.
When every individual elements are Prandtl-Reuss models, the
corresponding multi-layer model is indicated as of the
\emph{Prandtl-Reuss type}. This is the case in the present
section. Hence, let
\begin{equation}
\Psi =\sum_{i=1}^{N}\Psi _{i}=\sum_{i=1}^{N}\left[
\frac{1}{2}\left( \mbox{\boldmath $
\varepsilon-\varepsilon$}_{i}^{p}\right) :\mathbf{C}:\left(
 \mbox{\boldmath $\varepsilon
-\varepsilon$}_{i}^{p}\right) +\xi _{i}\left( \zeta _{i}\right)
\right]
\end{equation}
be the Helmholtz energy density, defined as the sum of $N$
expressions of the type (\ref{PsiClassIsoTwice}). The internal
variable $\mbox{\boldmath $\varepsilon$} _{i}^{p}$ is the plastic
strain of the generic element ${i}$, while $\zeta _{i}$ is the
scalar variable associated with the isotropic hardening of the
same element. All elements have by definition the same elastic
modulus tensor, chosen to be equal to
$\mathbf{C=}\frac{1}{N}\left[ \left( K-\frac{ 2}{3}G\right)
\mathbf{1\otimes 1+}2G\mathbf{I}\right] $. The non-dissipative
thermodynamic forces read:
\begin{equation}
\begin{array}{c}
\mbox{\boldmath $\sigma$}^{nd}=\sum_{i=1}^{N}\mathbf{C}:\left(
\mbox{\boldmath $\varepsilon$} - \mbox{\boldmath
$\varepsilon$}_{i}^{p} \right) , \ \ \ \mbox{\boldmath $\tau$}
_{i}^{nd}=-\mathbf{C}:\left(\mbox{\boldmath
$\varepsilon-\varepsilon$} _{i}^{p}\right) , \ \ \
R_{i}^{nd}=\frac{d\xi _{i}\left( \zeta _{i}\right)}{d\zeta_{i}}
\end{array}
\label{multiLayerndiss}
\end{equation}
Let us introduce the pseudo-potential $\phi$ as the sum of $N$
\emph{independent} functions of the type (\ref {PseudoReussIso1}):
\begin{equation}
\begin{array}{l}
\phi=\sum_{i=1}^{N}\phi _{i}\left( \mbox{\boldmath
$\dot{\varepsilon}$}^{\prime},\mbox{\boldmath
$\dot{\varepsilon}$}_{i}^{p^{\prime}},\dot{
\zeta}_{i}^{\prime};\zeta_{i}\right) \\
\ \ \ =\sum_{i=1}^{N}\left[\left( \sqrt{\frac{2}{3}}\sigma _{yi} \
g_{i}\left( \zeta _{i}\right) -\frac{d\xi _{i}\left( \zeta
_{i}\right)}{d\zeta_{i}} \right)
\dot{\zeta}_{i}^{\prime}+\mathbb{I}_{\bar{\mathbb{D}}_{i}}\left(
\mbox{\boldmath $\dot{\varepsilon}$}^{\prime},\mbox{\boldmath
$\dot{\varepsilon}$}_{i}^{p^{\prime}},\dot{\zeta}_{i}^{\prime}\right)\right]  \\
\\
\ \bar{\mathbb{D}}_{i}=\left\{
\begin{array}{l}
\left( \mbox{\boldmath
$\dot{\varepsilon}$}^{\prime},\mbox{\boldmath
$\dot{\varepsilon}$}_{i}^{p^{\prime}},\dot{
\zeta}_{i}^{\prime}\right) \in \mathbb{V}\textrm{ \ \ such \ that
\ }
 tr\left( \mbox{\boldmath $\dot{\varepsilon}$}_{i}^{p^{\prime}}\right) =0  \textrm{
\ and \ }\dot{\zeta}_{i}^{\prime}\geq \left\Vert \mbox{\boldmath
$\dot{\varepsilon}$} _{i}^{p^{\prime}}\right\Vert
\end{array}
\right\}
\end{array}
\end{equation}
The limit stresses $\sigma _{yi}$ as well as the isotropic
hardening functions $g_{i}\left( \zeta _{i}\right) $ are, in
general, distinct. The conjugated pseudo-potential is in turn the
sum of $N$ independent functions, i.e.
$\phi^{\ast}=\sum_{i=1}^{N}\phi^{\ast}_{i}$ with
\begin{equation}
\begin{array}{l}
\phi _{i}^{\ast }\left( \mbox{\boldmath $\sigma
$}^{d^{\prime}},\mbox{\boldmath $\tau
$}_{i}^{d^{\prime}},R_{i}^{d^{\prime}}\right)=\sup_{\left(
{\mbox{\scriptsize\boldmath
$\dot{\varepsilon}$}^{\prime},\mbox{\scriptsize\boldmath
$\dot{\varepsilon}$}}
_{i}^{p^{\prime}},\dot{\zeta}_{i}^{\prime}\right) \mathbf{\in
}\bar{\mathbb{D}}_{i}}\left( \mbox{\boldmath $
\sigma$}^{d^{\prime}}:\mbox{\boldmath
$\dot{\varepsilon}$}^{\prime}+\mbox{\boldmath $\tau
$}_{i}^{d^{\prime}}:\mbox{\boldmath $\dot{
\varepsilon}$}_{i}^{p^{\prime}}+R_{i}^{d^{\prime}}\dot{\zeta}_{i}^{\prime}-\phi _{i}\right)  \\
\textrm{ \ \ \ \ \ }=\mathbb{I}_{0}\left( \mbox{\boldmath $\sigma
$}^{d^{\prime}}\right) +\mathbb{ I}_{\mathbb{E}_{i}}\left(
\mbox{\boldmath $\tau
$}_{i}^{d^{\prime}},R_{i}^{d^{\prime}}\right)
\end{array}
\end{equation}
where $\mathbb{E}_{i}=\left\{ \left( \mbox{\boldmath
$\tau$}_{i}^{d^{\prime}},R_{i}^{d^{\prime}}\right) \in
\mathbb{S}^{2}\mathbb{\times R}\textrm{ \ \ such \ that \
}f_{i}\left(\mbox{\boldmath
$\tau$}_{i}^{d^{\prime}},R_{i}^{d^{\prime}};\zeta_{i}\right) \leq
0 \right\} $ and
\begin{equation}
f_{i}\left(\mbox{\boldmath
$\tau$}_{i}^{d^{\prime}},R_{i}^{d^{\prime}};\zeta_{i}\right)=\left\Vert
dev\left( \mbox{\boldmath $\tau$}_{i}^{d^{\prime}}\right)
\right\Vert -\sqrt{ \frac{2}{3}}\sigma _{yi} \ g_{i}\left( \zeta
_{i}\right) +R_{i}^{d^{\prime}}+\frac{d\xi _{i}\left( \zeta
_{i}\right)}{d\zeta_{i}}
\end{equation}
Therefore, $N$ independent loading surfaces have been defined.
Using the standard procedure based on the normality assumption,
$N$ pairs of flow rules of the type (\ref{flowReuss}) can be
derived:
\begin{equation}
\begin{array}{l}
\mbox{\boldmath $\dot{\varepsilon}$}_{i}^{p}=\frac{dev\left(
\mbox{\footnotesize\boldmath $\tau$}_{i}^{d}\right) }{\left\Vert
dev\left(\mbox{\footnotesize\boldmath $\tau$}_{i}^{d}\right)
\right\Vert } \dot{\lambda}_{i}=
\mathbf{n}_{i}\dot{\lambda}_{i}, \ \ \ \ \ \ \dot{\zeta}_{i}=\dot{\lambda}_{i} \\
\textrm{with \ \ \ }\dot{\lambda}_{i}\geq 0, \ \ \ f_{i}\leq 0, \
\ \ \ \dot{\lambda}_{i}f_{i}=0
\end{array}
\label{flowReussmulti}
\end{equation}
Moreover, by imposing the consistency conditions and accounting
for Eqs. (\ref{multiLayerndiss}) as well as the identities
$\mbox{\boldmath $\tau$}_{i}^{d}=-\mbox{\boldmath
$\tau$}_{i}^{nd}$ and $ R_{i}^{d}=-R_{i}^{nd}$, each plastic
multiplier can be easily determined by an expression of the type
(\ref{lambdaReussMod}):
\begin{equation}
\dot{\lambda}_{i}=H\left( f_{i}\right) \frac{\left\langle
\mathbf{n}_{i}:\mathbf{C}:\mbox{\boldmath
$\dot{\varepsilon}$}\right\rangle
}{\mathbf{n}_{i}:\mathbf{C:n}_{i}+\sqrt{\frac{2}{3}}\sigma
_{yi}\frac{dg_{i}\left( \zeta_{i} \right)}{d\zeta_{i}} }=H\left(
f_{i}\right) \frac{\left\langle \mathbf{n}_{i}:\mbox{\boldmath
$\dot{\varepsilon}$}\right\rangle} {1\mathbf{+}\sqrt{\frac{2}{3}}
\frac{\sigma_{yi}}{2G}\frac{dg_{i}\left( \zeta_{i}
\right)}{d\zeta_{i}} } \label{lambdaReussmulti}
\end{equation}
provided that $1+\sqrt{\frac{2}{3}}\frac{\sigma
_{yi}}{2G}\frac{dg_{i}\left( \zeta_{i} \right)}{d\zeta_{i}} >0$.

The Distributed Element Model \citep {Iwan66} \citep{Chiang94a} is
recovered when $g_{i}\left( \zeta _{i}\right) =1 $ and $\xi
_{i}\left( \zeta _{i}\right) =0$.

\section{Endochronic theory}

Endochronic theory was first formulated by \cite{Valanis71}, who
suggested the use of a positive scalar variable $\vartheta $,
called \emph{ intrinsic time}, in the definition of constitutive
laws of plasticity models. The evolution laws are described by
convolution integrals involving past values of the state variable
$\mbox{\boldmath $\varepsilon$}$ and suitable scalar functions
depending on $\vartheta $ called \emph{memory kernels}. When the
memory kernel is exponential, the integral expressions can be
rewritten as simple differential equations, which, for an
initially isotropic endochronic material fulfilling the plastic
incompressibility assumption, read:
\begin{equation}
\left\{
\begin{array}{l}
tr\left( \mbox{\boldmath $\dot{\sigma}$}\right)
=3K \ tr\left( \mathbf{\mbox{\boldmath $\dot{\varepsilon}$}}\right)  \\
dev\left( \mbox{\boldmath $\dot{\sigma}$}\right) =2G \ dev\left(
\mathbf{\mbox{\boldmath $\dot{\varepsilon}$}}\right)
\mathbf{-}\beta \ dev\left( \mbox{\boldmath $\sigma$} \right) \
\dot{\vartheta}
\end{array}
\right. \label{EndoFormulGen}
\end{equation}
with $\beta >0$. These relationships are equivalent to:
\begin{equation}
\left\{
\begin{array}{l}
\mbox{\boldmath$\sigma$}\mathbf{=C:}\left( \mbox{\boldmath
$\varepsilon-\varepsilon$}^{p}\right)
\textrm{, } \\
\mathbf{C=}\left( K-\frac{2}{3}G\right) \mathbf{1\otimes
1+}2G\mathbf{I}
\textrm{ ,} \\
tr\left( \mbox{\boldmath $\dot{\varepsilon}$}^{p}\right)
=0\textrm{ \ \ \ \ \ \ \ and \ \ \ \ \ \ \ \ \ }\mbox{\boldmath
$\dot{\varepsilon}$}^{p}=\frac{dev\left(\mbox{\footnotesize\boldmath
$\sigma$}\right) }{2G/\beta } \ \dot{\vartheta}
\end{array}
\right. \label{endoGen1}
\end{equation}
where $\dot{\vartheta}\geq 0$ is the time-derivative of the
intrinsic time. The simplest choice for the intrinsic time flow is
$\dot{\vartheta}=\left\Vert dev\left( \mbox{\boldmath
$\dot{\varepsilon}$}\right) \right\Vert $ \citep{Valanis71}.
However, more complex definitions can be given, such as:
\begin{equation}
\begin{array}{c}
\dot{\vartheta}=\frac{\dot{\zeta}}{g\left( \zeta \right)
}=f_{1}\left( \zeta \right) \dot{\zeta}\textrm{ \ \ \ \ \ \ \ \
with \ \ \ }\dot{\zeta}=\left\Vert dev\left( \mbox{\boldmath
$\dot{\varepsilon}$}\right) \right\Vert
\end{array}
\label{IntrTimeDegr}
\end{equation}
where $\zeta $ is the \emph{intrinsic time scale} and the positive
function $f_{1}(\zeta)=1/g(\zeta)$, such that $f_{1}(0)=1$, is
sometimes called hardening-softening function \citep{Bazant76}.

\subsection{A new formulation of endochronic models}

In this section, the endochronic model defined by Eqs.
(\ref{endoGen1}) is innovatively described by its Helmholtz free
energy and a suitable pseudo-potential associated with generalized
normality conditions. This approach allows for insightful
comparisons between endochronic models and Prandtl-Reuss models.
The main implications will be discussed later. Let
$\mathbf{v=}\left( \mbox{\boldmath$\varepsilon$},\mbox{\boldmath
$\varepsilon$}^{p},\zeta \right) $ and
$\mathbf{q}^{nd}\mathbf{=}\left( \mbox{\boldmath $\sigma$}
^{nd},\mbox{\boldmath $\tau$}^{nd},R^{nd}\right) $ be the assumed
state variables and the associated non-dissipative thermodynamic
forces, respectively. They are the same as in the Prandtl-Reuss
model with isotropic hardening. The Helmholtz free energy $\Psi $
reads:
\begin{equation}
\Psi =\frac{1}{2}\left( \mbox{\boldmath
$\varepsilon-\varepsilon$}^{p}\right) : \mathbf{C}:\left(
\mbox{\boldmath $\varepsilon-\varepsilon$}^{p}\right)
\label{PsiEndoIso}
\end{equation}
This form is a particular case of the one originally proposed by
\cite{Valanis71}, since only one tensorial internal variable, the
plastic strain, is considered here. The first two non-dissipative
forces $\mbox{\boldmath $\sigma$} ^{nd}$ and $\mbox{\boldmath
$\tau$}^{nd}$ are the same as in Eq. (\ref{J2isoNonDissip}), while
$R^{nd}=0 $ since $\Psi $ is assumed to be independent of the
scalar variable $\zeta $. The pseudo-potential is defined as
follows:
\begin{equation}
\begin{array}{l}
\phi\left(\mbox{\boldmath
$\dot{\varepsilon}$}^{\prime},\mbox{\boldmath$\dot{\varepsilon}$}^{p^{\prime}}
,\dot{\zeta}^{\prime};\mbox{\boldmath
$\varepsilon$},\mbox{\boldmath $\varepsilon$}^{p},\zeta \right)
=\frac{\left\Vert dev\left[
\mathbf{C}:\left(\mbox{\footnotesize\boldmath
$\varepsilon-\varepsilon$}^{p}\right) \right] \right\Vert
^{2}}{2Gg\left( \zeta \right) /\beta
}\dot{\zeta}^{\prime}+\mathbb{I}_{\bar{\mathbb{D}}}\left(\mbox{\boldmath
$\dot{\varepsilon}$}^{\prime},\mbox{\boldmath$\dot{\varepsilon}$}^{p^{\prime}}
,\dot{\zeta}^{\prime};\mbox{\boldmath
$\varepsilon$},\mbox{\boldmath $\varepsilon$}^{p},\zeta \right) \\
 \\
\bar{\mathbb{D}}=\left\{
\begin{array}{l}
\left( \mbox{\boldmath$\dot{\varepsilon}$}^{\prime},
\mbox{\boldmath$\dot{\varepsilon}$}^{p^{\prime}},\dot{\zeta}^{\prime}\right)
\in \mathbb{V}\textrm{ \ \ \ \ \ such \ that} \\
\ tr\left( \mbox{\boldmath
$\dot{\varepsilon}$}^{p^{\prime}}\right) =0 \textrm{ , \ \ \ \
}\mbox{\boldmath $\dot{\varepsilon}$}^{p^{\prime}}=\frac{dev\left[
\mathbf{C}:\left( \mbox{\footnotesize\boldmath
$\varepsilon-\varepsilon$}^{p}\right) \right] }{ \frac{2G}{\beta }
\ g\left( \zeta \right) } \ \dot{\zeta}^{\prime}\textrm{ ,}
 \ \ \ \ \dot{\zeta}^{\prime}\geq 0
\end{array}
\right\}
\end{array}
\label{fiEndoIso}
\end{equation}
The first term of $\phi$, in which the stress $\mbox{\boldmath
$\sigma$}^{nd}=\mathbf{C}:\left( \mbox{\boldmath
$\varepsilon-\varepsilon$}^{p}\right) $ is written as a function
of the state variables, is equal to the intrinsic dissipation
$\Phi _{m}$ when $\dot{\zeta}^{\prime}$ assumes the actual value
$\dot{\zeta}$. The first condition associated with the closed
convex set $\bar{\mathbb{D}}$ introduces the plastic
incompressibility assumption, while the second condition
characterizes the plastic strain flow of endochronic theory, as it
can be seen by comparing it to Eqs. (\ref {endoGen1}) and
(\ref{IntrTimeDegr}). Finally, the positivity of
$\dot{\zeta}^{\prime} $ is imposed in order to guarantee that
$\phi $ is positive. Using the language of the endochronic theory,
the internal variable $\zeta $ corresponds to the intrinsic time
scale, while the intrinsic time $\vartheta $ is defined by its
flow $\dot{ \vartheta}=\dot{\zeta}/g\left( \zeta \right).$ The
variable $ \zeta $ does not directly appear in the Helmholtz free
energy density and its associated thermodynamic forces,
dissipative and non-dissipative, are thus zero. However, $\zeta $
is not zero during the plastic evolution and plays an important
role in the definition of $\mbox{\boldmath
$\dot{\varepsilon}$}^{p}$.

\begin{figure}
\begin{center}
\includegraphics[width=13.4cm]{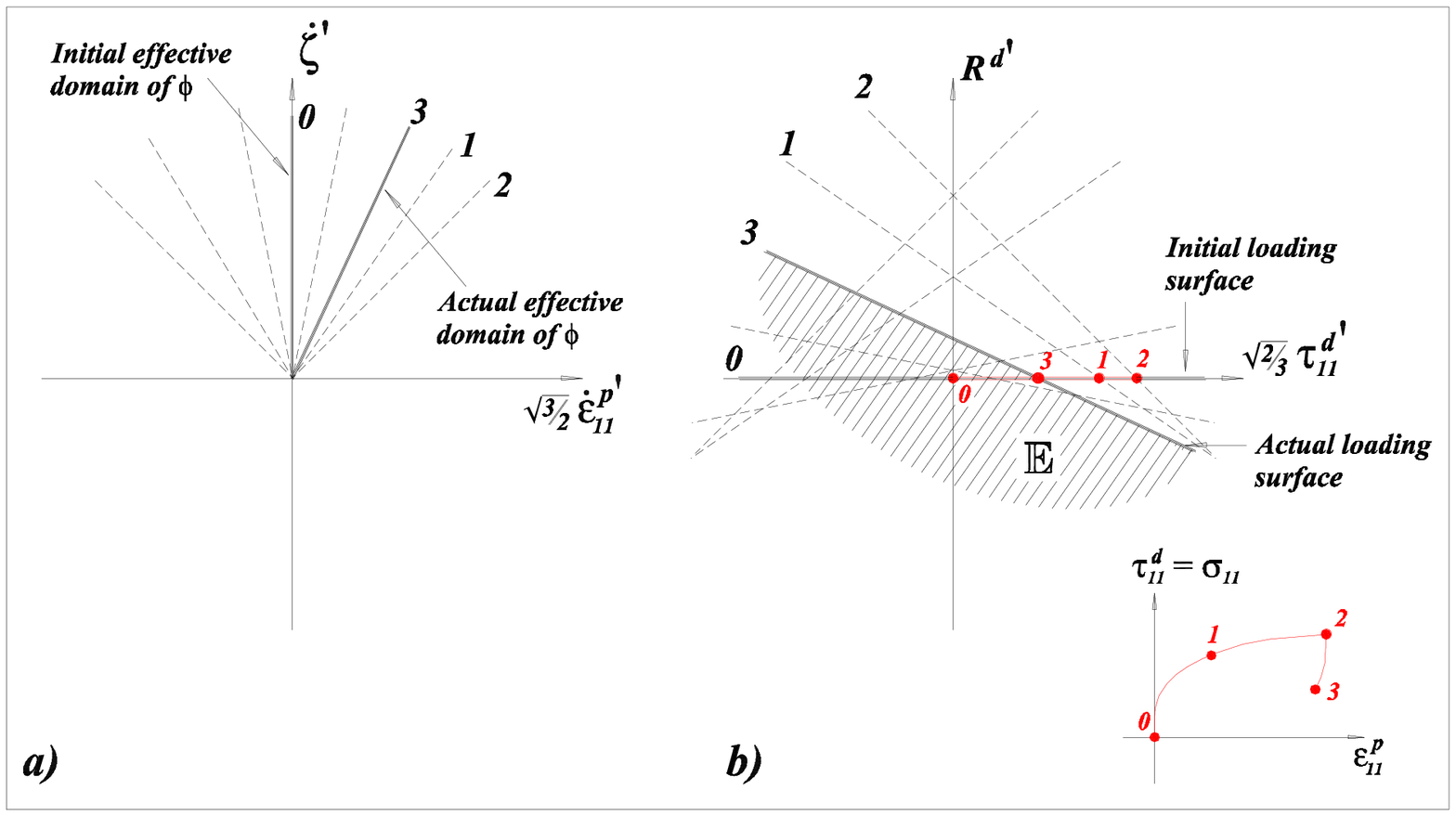}
\caption{Endochronic model. Tension-compression case with
$g(\zeta)=1$. \newline a) Several configurations of the set
$\mathbb{D}$, which is the projection of the pseudo-potential
effective domain $\bar{\mathbb{D}}$ on the $\left(\mbox{\boldmath
$\dot{\varepsilon}$}^{p^{\prime}},\dot{\zeta}^{\prime}\right)$-plane.
b) Configurations of the convex set $\mathbb{E}$ associated with
those of $\mathbb{D}$. The point $(\mbox{\boldmath
$\tau$}^{d},R^{d})$, representing the actual state, always lies on
the axis $R^{d^{\prime}}=0$.}

\label{FigEndo}
\end{center}
\end{figure}

The conjugated pseudo-potential is, in this case, of the following
form:
\begin{equation}
\begin{array}{l}
\phi ^{\ast }\left(\mbox{\boldmath $\sigma$}^{d^{\prime}},
\mbox{\boldmath
$\tau$}^{d^{\prime}},R^{d^{\prime}};\mbox{\boldmath
$\varepsilon$},\mbox{\boldmath $\varepsilon$}^{p},\zeta \right)=
\\
\ \ \ \ =\sup_{\left( \mbox{\scriptsize\boldmath
$\dot{\varepsilon}$}^{\prime},\mbox{\scriptsize\boldmath
$\dot{\varepsilon}$}^{p^{\prime}}, \dot{\zeta}^{\prime}\right)
\mathbf{\in }\bar{\mathbb{D}}}\left( \mbox{\boldmath
$\sigma$}^{d^{\prime}}: \mbox{\boldmath
$\dot{\varepsilon}$}^{\prime}+\mbox{\boldmath
$\tau$}^{d^{\prime}}:\mbox{\boldmath $\dot{\varepsilon}$}
^{p^{\prime}}+R^{d^{\prime}}\dot{\zeta}^{\prime}-\phi \right) \\ \
\ \ \ =\mathbb{I} _{0}\left( \mbox{\boldmath
$\sigma$}^{d^{\prime}}\right)+\mathbb{I}_{\mathbb{E}} \left(
\mbox{\boldmath
$\tau$}^{d^{\prime}},R^{d^{\prime}};\mbox{\boldmath
$\varepsilon$},\mbox{\boldmath $\varepsilon$}^{p},\zeta
\right)\end{array}\label{pseudoEndoIso}
\end{equation}
where $\mathbb{E}=\left\{ \left( \mbox{\boldmath
$\tau$}^{d^{\prime}},R^{d^{\prime}}\right) \in \mathbb{S}^{2}
\mathbb{\times R}\textrm{ \ \ such \ that \ }f\left(
\mbox{\boldmath
$\tau$}^{d^{\prime}},R^{d^{\prime}};\mbox{\boldmath
$\varepsilon$},\mbox{\boldmath $\varepsilon$}^{p},\zeta\right)\leq
0\right\}$  and
\begin{equation}
\begin{array}{l}
f \left( \mbox{\boldmath
$\tau$}^{d^{\prime}},R^{d^{\prime}};\mbox{\boldmath
$\varepsilon$},\mbox{\boldmath
$\varepsilon$}^{p},\zeta\right)=\frac{dev\left(
\footnotesize{\mbox{\boldmath $\tau$}^{d^{\prime}}}\right)
:dev\left[ \mathbf{C}:\left( \footnotesize{\mbox{\boldmath
$\varepsilon-\varepsilon$}}^{p}\right) \right] }{2Gg\left( \zeta
\right) /\beta }-\frac{\left\Vert dev\left[ \mathbf{C}:\left(
\footnotesize{\mbox{\boldmath $ \varepsilon -\varepsilon
$}}^{p}\right) \right] \right\Vert ^{2}}{2Gg\left( \zeta \right)
/\beta }+R^{d^{\prime}}
\end{array} \label{fEndoIso}
\end{equation}
The expression (\ref{fEndoIso}) defines the \emph{loading function
of endochronic models}. It is associated with a set $\mathbb{E}$
in the $\left( \mbox{\boldmath $\tau
$}^{d^{\prime}},R^{d^{\prime}}\right)$ space. In Figure
\ref{FigEndo} this set is represented in the case of
tension-compression with $g\left( \zeta \right) =1$, together with
the projection of $\bar{\mathbb{D}}$ on the $\left(\mbox{\boldmath
$\dot{\varepsilon}$}^{p^{\prime}},\dot{\zeta}^{\prime}
\right)$-plane. This last set is indicated by $\mathbb{D}$. Some
important remarks have to be made. First, as the system evolves,
both sets change, due to their dependence on the internal
variables. At every instantaneous configurations, the set
$\mathbb{D}$ is a straight line starting from the origin. The
corresponding sets $\mathbb{E}$ are half-planes orthogonal to $
\mathbb{D}$. Moreover, Eq. (\ref{PsiEndoIso}) entails that
$R^{nd}=-R^{d}=0$ and, accounting for the indicator function
$\mathbb{I}_{0}(\mbox{\boldmath $\sigma$}^{d^{\prime}})$ in
(\ref{pseudoEndoIso}), it also leads to
\begin{equation}
\mbox{\boldmath $\tau$}^{d}=-\mbox{\boldmath
$\tau$}^{nd}=\mbox{\boldmath $\sigma$}\mathbf{=C}:\left(
\mbox{\boldmath $ \varepsilon -\varepsilon $}^{p}\right)
\label{taundEps}
\end{equation}
Therefore, at the actual stress state $(\mbox{\boldmath
$\tau$}^{d},R^{d})$ the loading function $f$ \emph{is always equal
to zero}. In other words, $(\mbox{\boldmath $\tau$}^{d},R^{d})$
always belongs to $\partial \mathbb{E}$, during both loading and
unloading phases, and \emph{all the states are plastic states}.
The normality conditions lead to the endochronic flow rules:
\begin{equation}
\begin{array}{c}
\mbox{\footnotesize\boldmath
$\dot{\varepsilon}$}^{p}=\frac{dev\left[
\mathbf{C}:\left(\mbox{\footnotesize\boldmath $\varepsilon
-\varepsilon$}^{p}\right) \right] }{2G \ g\left( \zeta \right)
/\beta }\dot{\lambda}, \ \ \ \ \ \ \ \dot{\zeta}=\dot{\lambda}
\textrm{ \ \ \ \ \ \ \ \ \ with\ \ }\dot{\lambda}\geq 0
\end{array}
\label{flowendoIso}
\end{equation}
Eqs. (\ref{pseudoEndoIso})-(\ref{fEndoIso}) and
(\ref{flowendoIso}) prove that endochronic models are associative
in generalized sense. Moreover, since $f$ is always equal to zero
at the actual state, the loading-unloading conditions reduce to
the requirement of the plastic multiplier $\dot{\lambda}$ to be
non-negative (see the inequality in (\ref{flowendoIso})). In
addition, the time derivative $\dot{f}$ at $(\mbox{\boldmath
$\tau$}^{d},R^{d})$, computed accounting for the fact that $f$
also depends on $\mbox{\boldmath $\varepsilon$}$, $\mbox{\boldmath
$\varepsilon$}^{p}$ and $\zeta$, is also equal to zero and
therefore, \emph{the consistency condition is automatically
fulfilled and cannot be used to compute }$\dot{\lambda}$.

\begin{figure}
\begin{center}
\includegraphics[width=13.4cm]{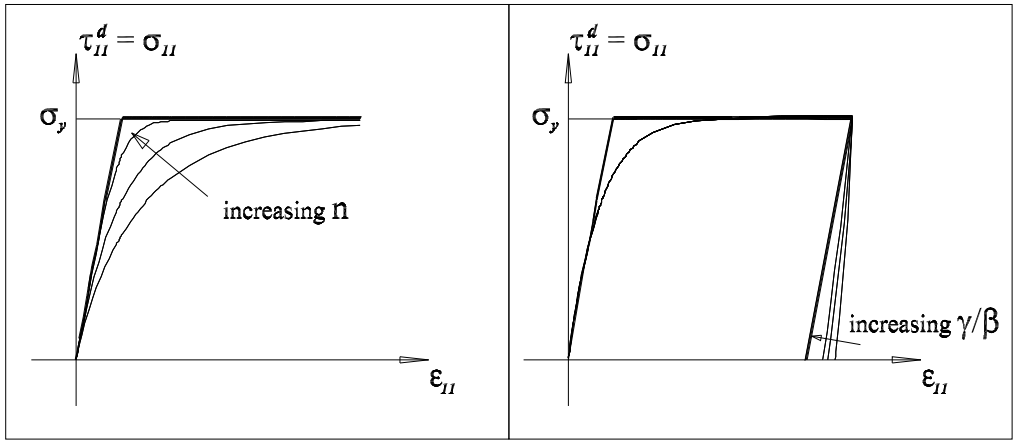}
\caption{Endochronic Karray-Bouc-Casciati model (thin lines) vs.
Prandtl-Reuss model (thick line). Tension-compression case with
$g(\zeta)$=1 and $\xi(\zeta)=0$. \newline a) Influence of the
parameter $n$ on loading branches. b) Influence of the
$\gamma/\beta$ ratio on unloading branches. The slope at
$\sigma_{11}=0$ is the same for all $\gamma/\beta$ values.}

\label{FigEndoPrandtlSigEps}
\end{center}
\end{figure}

This situation is typical of endochronic theory and entails that
the plastic multiplier $\dot{\lambda}=\dot{\zeta}$ has to be
defined by an additional assumption. When the function $g\left(
\zeta \right) $ is also fixed, the plastic flow $\mbox{\boldmath
$\dot{\varepsilon}$}^{p}$ and the intrinsic time flow
$\dot{\vartheta}=\dot{\zeta} / g\left( \zeta \right)$\ are then
known. The standard choices are $g\left( \zeta \right) =1$ and
$\dot{ \vartheta}=\dot{\zeta}=\left\Vert dev\left( \mbox{\boldmath
$\dot{\varepsilon}$} \right) \right\Vert $. It has been shown in
\cite {Erlicher04} that more complex definitions can be chosen,
such as $g\left( \zeta \right) =1$ and
\begin{equation}
\begin{array}{l}
\dot{\vartheta}=\dot{\zeta}=\left\Vert dev\left( \mbox{\boldmath
$\tau$}^{d}\right) \right\Vert ^{n-2}\left( 1+\frac{\gamma }{\beta
}sign\left( dev\left( \mbox{\boldmath $\tau$}^{d}\right)
:\mbox{\boldmath $\dot{\varepsilon}$}\right) \right) \ | dev\left(
\mbox{\boldmath $\tau$}^{d}\right) :\mbox{\boldmath
$\dot{\varepsilon}$}
|  \\
-\beta \leq \gamma \leq \beta \ , \ \ \ \ n>0
\end{array}
\label{BoucWen}
\end{equation}
which is effectively the Karray-Bouc-Casciati model
\citep{Karray89} \citep {Casciati89}. It must be noticed that both
flows $\mbox{\boldmath $\dot{\varepsilon}$}^{p}$ and $\dot{\zeta}$
can be different from zero during unloading phases, i.e. when
$dev\left( \mbox{\boldmath $\tau$} ^{d}\right) :\mbox{\boldmath
$\dot{\varepsilon}$}<0$. This situation, which is not possible in
classical plasticity, occurs when $\gamma \neq \beta$.  Figure
\ref{FigEndoPrandtlSigEps}a illustrates for the mono-dimensional
case the effect of $n$ for given values of the other parameters:
in the limit of increasing $n$-values the Prandtl-Reuss model is
retrieved. Figure \ref{FigEndoPrandtlSigEps}b shows unloading
branches for different $\gamma/\beta$ ratios, the other parameters
being fixed: plastic strains may occur and tend to zero when
$\gamma/\beta$ tends to $1$.

\subsection{Endochronic theory vs. Prandtl-Reuss model}

\begin{figure}
\begin{center}
\includegraphics[width=13.4cm]{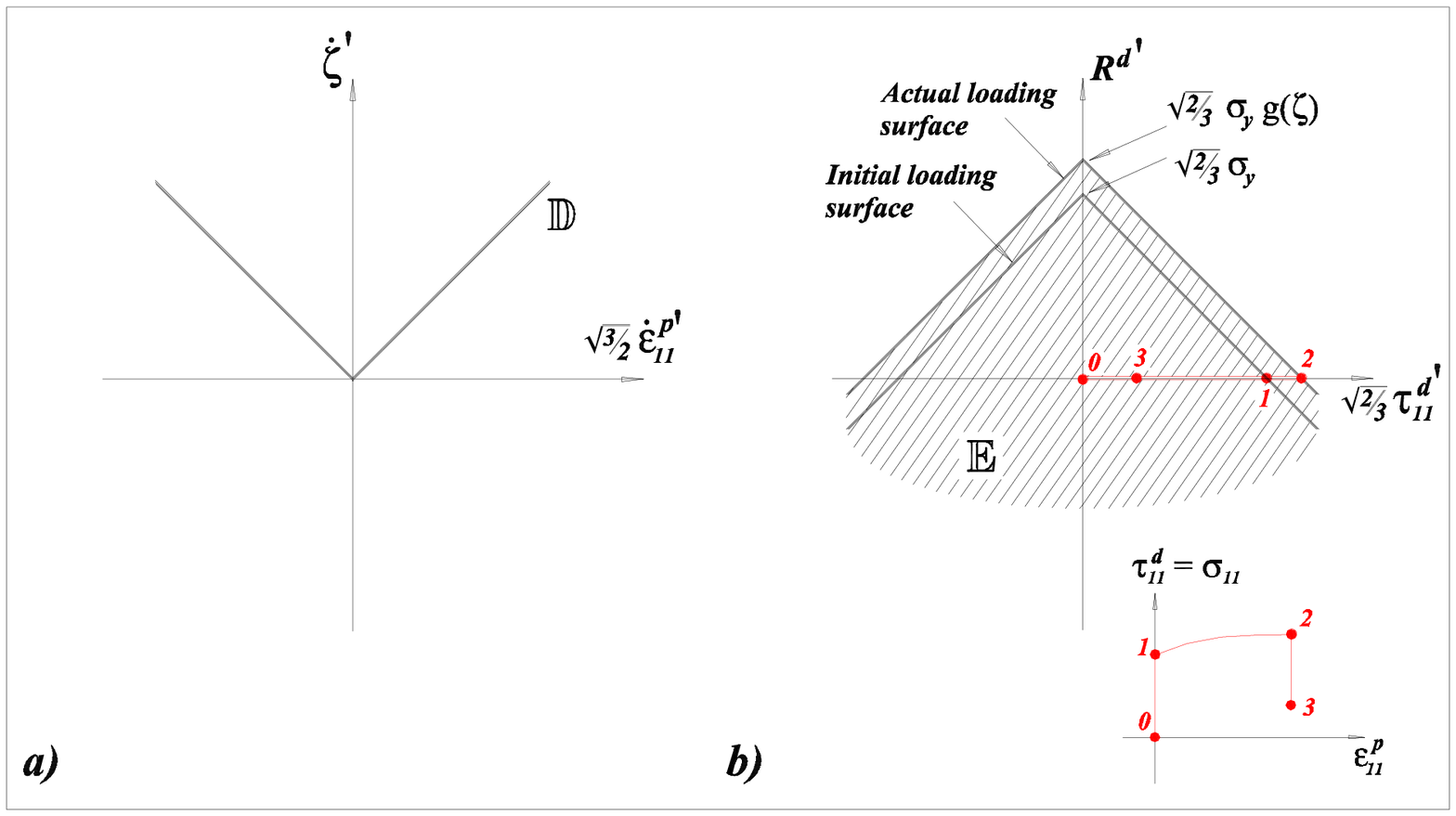}
\caption{Endochronic model vs. Prandtl-Reuss model.
Tension-compression case. \newline a) The set $\mathbb{D}$ is the
projection of $\bar{\mathbb{D}}$ on the $\left(\mbox{\boldmath
$\dot{\varepsilon}$}^{p^{\prime}},\dot{\zeta}^{\prime}\right)$-plane,
where $\bar{\mathbb{D}}$ is the non-convex effective domain of the
pseudo-potential $\tilde{\phi}$ of Eq. (\ref{fiClassEndo}). It
defines an endochronic model where the intrinsic time flow
$\dot{\zeta}$ equals the norm of $\mbox{\boldmath
$\dot{\varepsilon}$}^{p}$. b) The convex set $\mathbb{E}$
associated with the indicator function $\phi^{\ast}$ given in Eqs.
(\ref{fistar1})-(\ref{fistar2}), which is the Legendre-Fenchel
conjugated of $\tilde{\phi}$.}

\label{FigEndoPrandtl}
\end{center}
\end{figure}

Consider the endochronic model, as formulated in the previous
section, and the \emph{modified} Prandtl-Reuss model. The
significant state variables $\mbox{\boldmath $\varepsilon$},
\mbox{\boldmath $\varepsilon$}^{p}$ and $\zeta $ are the same in
both cases. Moreover, Eqs. (\ref{PsiClassIsoTwice}) and
(\ref{PsiEndoIso}) show that the Helmholtz free energies differ
only by the term $\xi \left( \zeta \right) $, which is zero in
endochronic theory. The main differences concern
pseudo-potentials, as seen comparing Eqs. (\ref{PseudoReussIso1})
and (\ref{fiEndoIso}). However, the strict relationship between
the two models can be highlighted by imposing that $\dot{\zeta
}^{\prime}=\left\Vert \mbox{\boldmath
$\dot{\varepsilon}$}^{p^{\prime}}\right\Vert $ in
(\ref{fiEndoIso}): when $\dot{\zeta}^{\prime}>0$, the condition
$\left\Vert dev\left(\mathbf{C:}\left( \mbox{\boldmath
$\varepsilon-\varepsilon$} ^{p}\right) \right) \right\Vert
=\frac{2G}{\beta }g\left( \zeta \right) $ must be fulfilled, while
for $\dot{\zeta}^{\prime}=0$ there is no limitation on
$dev\left(\mathbf{C:}\left(\mbox{\boldmath
$\varepsilon-\varepsilon$}^{p}\right)\right) $. As a result, the
endochronic pseudo-potential (\ref {fiEndoIso}) becomes equal to
\begin{equation}
\begin{array}{l}
\tilde{\phi}\left( \mbox{\boldmath
$\dot{\varepsilon}$}^{\prime},\mbox{\boldmath
$\dot{\varepsilon}$}^{p^{\prime}},\dot{\zeta}^{\prime};\zeta\right)=\frac{2G}{\beta
} \ g\left( \zeta \right) \ \dot{
\zeta}^{\prime}+\mathbb{I}_{\bar{\mathbb{D}}}\left(
\mbox{\boldmath $\dot{\varepsilon}$}^{\prime},
\mbox{\boldmath $\dot{\varepsilon}$}^{p^{\prime}},\dot{\zeta}^{\prime}\right)  \\
\ \bar{\mathbb{D}}\mathbf{=}\left\{
\begin{array}{l}
\left( \mbox{\boldmath
$\dot{\varepsilon}$}^{\prime},\mbox{\boldmath
$\dot{\varepsilon}$}^{p^{\prime}},\dot{\zeta}^{\prime}\right) \in
\mathbb{V}\textrm{ \ such \ that }  \ tr\left( \mbox{\boldmath
$\dot{\varepsilon}$}^{p^{\prime}}\right) =0\textrm{ \ \ and \ \
}\dot{\zeta}^{\prime}=\left\Vert \mbox{\boldmath
$\dot{\varepsilon}$}^{p^{\prime}}\right\Vert
\end{array}
\right\}
\end{array}
\label{fiClassEndo}
\end{equation}
The set $\bar{\mathbb{D}}$ and the function $\tilde{\phi}$ are not
convex (see Figure \ref{FigEndoPrandtl}a). However, the
Legendre-Fenchel conjugate of $\tilde{\phi}$ is still well-posed
(Appendix, item 5) and can be explicitly derived from the standard
procedure:
\begin{equation}
\begin{array}{l}
\phi ^{\ast }\left(\mbox{\boldmath $\sigma$}^{d^{\prime}},
\mbox{\boldmath $\tau$}^{d^{\prime}},R^{d^{\prime}};\zeta \right)
\\
\ \ \ \ \ \ =\sup_{\left( \mbox{\scriptsize\boldmath
$\dot{\varepsilon}$}^{\prime},\mbox{\scriptsize\boldmath
$\dot{\varepsilon}$}^{p^{\prime}}, \dot{\zeta}^{\prime}\right)
\mathbf{\in }\bar{\mathbb{D}}}\left( \mbox{\boldmath
$\sigma$}^{d^{\prime}}: \mbox{\boldmath
$\dot{\varepsilon}$}^{\prime}+\mbox{\boldmath
$\tau$}^{d^{\prime}}:\mbox{\boldmath $\dot{\varepsilon}$}
^{p^{\prime}}+R^{d^{\prime}}\dot{\zeta}^{\prime}-\tilde{\phi} \right) \\
\ \ \ \ \ \ =\mathbb{I}_{0}\left( \mbox{\boldmath
$\sigma$}^{d^{\prime}}\right) +\mathbb{I}_{\mathbb{E}}\left(
\mbox{\boldmath $\tau$}^{d^{\prime}},R^{d^{\prime}};\zeta \right)
 \label{fistar1}
 \end{array}
\end{equation}
with $\mathbb{E}=\left\{ \left( \mbox{\boldmath
$\tau$}^{d^{\prime}},R^{d^{\prime}}\right) \in \mathbb{S
}^{2}\mathbb{\times R}\textrm{\ \ such \ that \ \ }f\left(
\mbox{\boldmath $\tau$}^{d^{\prime}},R^{d^{\prime}};\zeta
\right)\leq 0\right\} $ and
\begin{equation}
f\left( \mbox{\boldmath $\tau$}^{d^{\prime}},R^{d^{\prime}};\zeta
\right)=\left\Vert dev\left( \mbox{\boldmath
$\tau$}^{d^{\prime}}\right) \right\Vert -\frac{2G}{\beta }
 \ g\left( \zeta \right) +R^{d^{\prime}} \label{fistar2}
\end{equation}
Provided that $\frac{2G}{\beta }=\sqrt{\frac{2}{3}}\sigma _{y}$,
Eqs. (\ref{fistar1})-(\ref{fistar2}) also define the
Legendre-Fenchel conjugate of the proper convex lower
semi-continuous function (Appendix, item 5)
\begin{equation}
\begin{array}{l}
\phi =cl\left( conv \ \tilde{\phi}\right)
=\sqrt{\frac{2}{3}}\sigma _{y} \ g\left( \zeta \right) \
\dot{\zeta}^{\prime}+\mathbb{I}_{\bar{\mathbb{D}} }\left(
\mbox{\boldmath $\dot{\varepsilon}$}^{\prime},\mbox{\boldmath
$\dot{\varepsilon}$}^{p^{\prime}},\dot{\zeta}^{\prime}\right)
\\
\ \bar{\mathbb{D}}\mathbf{=}\left\{
\begin{array}{l}
\left( \mbox{\boldmath
$\dot{\varepsilon}$}^{\prime},\mbox{\boldmath
$\dot{\varepsilon}$}^{p^{\prime}},\dot{\zeta}^{\prime}\right) \in
\mathbb{V}\textrm{ \ \ \ such \ that}   \ \ \ tr\left(
\mbox{\boldmath $\dot{\varepsilon}$}^{p^{\prime}}\right)
=0\textrm{ \ \ \ and \ \ \ }\dot{\zeta}^{\prime}\geq \left\Vert
\mbox{\boldmath $\dot{\varepsilon}$}^{p^{\prime}}\right\Vert
\end{array}
\right\}
\end{array}
\label{fiClassEndo1}
\end{equation}
which corresponds to the pseudo-potential of a modified
Prandtl-Reuss model, in the case $\xi \left( \zeta \right)
=\frac{d\xi\left( \zeta \right)}{d\zeta} =0$ (see Eqs.
(\ref{PseudoReussIso1})).

A similar comparison between the classical Prandtl-Reuss model
(Eqs. (\ref{PsiClassIso}) and (\ref{PseudoReussIso})) and
endochronic models is possible as well, but only when the former
is perfectly plastic, i.e. if $\xi \left( \zeta \right) =0$, and
conditions $\xi \left( \zeta \right) =0$ and $g=1$ hold in the
latter. Note that these assumptions have been adopted in Figure
\ref{FigEndoPrandtlSigEps}.

\subsection{Multi-layer models of endochronic type \label{SecEndomulti}}

The concept of assembling in parallel several plastic elements can
be applied to the case in which each element is of endochronic
type. The approach is analogous to the one adopted in Section
\ref{SecmultiPrandtl}. Let $\mbox{\boldmath$\varepsilon$}$ and
$\left(\mbox{\boldmath$\varepsilon$}_{i}^{p},\zeta_{i}\right)$ be
the relevant state variables. Then, the Helmholtz energy is
defined as the sum of $N$ contributions, of the same kind as in
Eq. (\ref{PsiEndoIso}):
\begin{equation}
\Psi =\sum_{i=1}^{N}\Psi _{i}=\sum_{i=1}^{N}\left[
\frac{1}{2}\left( \mbox{\boldmath $ \varepsilon -\varepsilon
$}_{i}^{p}\right) :\mathbf{C}:\left( \mbox{\boldmath $
\varepsilon$} -\mbox{\boldmath $\varepsilon$}_{i}^{p}\right)
\right]
\end{equation}
where the internal variables
$\mbox{\boldmath$\varepsilon$}_{i}^{p}$ have the meaning of
plastic strain of the $i-th$ endochronic element. The
thermodynamic forces associated with $\zeta _{i}$\ are zero, viz.
$ R_{i}^{nd}=0 $. Moreover, $N$ \emph{independent}
pseudo-potentials are assumed to be of the type (\ref{fiEndoIso}):
\begin{equation}
\begin{array}{l}
\phi _{i} =\frac{\left\Vert dev\left[ \mathbf{C}:\left(
\mbox{\footnotesize\boldmath
$\varepsilon-\varepsilon$}_{i}^{p}\right) \right] \right\Vert
^{2}}{2G \ g_{i}\left( \zeta _{i}\right) /\beta _{i}} \
\dot{\zeta}_{i}^{\prime}+\mathbb{I} _{\bar{\mathbb{D}}_{i}}\left(
\mbox{\boldmath $\dot{\varepsilon}$}^{\prime},\mbox{\boldmath
$\dot{\varepsilon}$}
_{i}^{p^{\prime}},\dot{\zeta}_{i}^{\prime};\mbox{\boldmath
$\varepsilon$},\mbox{\boldmath $\varepsilon$}
_{i}^{p},\zeta_{i} \right) \\ \\
\bar{\mathbb{D}}_{i}=\left\{
\begin{array}{l}
\left( \mbox{\boldmath
$\dot{\varepsilon}$}^{\prime},\mbox{\boldmath
$\dot{\varepsilon}$}_{i}^{p^{\prime}},\dot{
\zeta}_{i}^{\prime}\right) \in \mathbb{V}\textrm{ \ \ \ \ \ such \ that } \\
tr\left( \mbox{\boldmath
$\dot{\varepsilon}$}_{i}^{p^{\prime}}\right) =0, \ \ \ \ \dot{
\zeta}_{i}^{\prime}\geq 0\textrm{ \ \ \  and \ \ \ }
\mbox{\boldmath$\dot{\varepsilon}$}_{i}^{p^{\prime}}=\frac{dev\left[
\mathbf{C}:\left(\mbox{\footnotesize\boldmath$\varepsilon
-\varepsilon$}_{i}^{p}\right) \right] }{2G \ g_{i}\left( \zeta
_{i}\right) /\beta _{i}} \ \dot{\zeta}_{i}^{\prime}
\end{array}
\right\}
\end{array}
\label{fimultiendo}
\end{equation}
with $\beta _{i}>0$, $g_{i}\left( \zeta _{i}\right) >0$ and
$g_{i}\left(0 \right) =1$. The pseudo-potential of the multi-layer
model is $\phi=\sum_{i=1}^{N}\phi_{i}$ and its dual is
$\phi^{*}=\sum_{i=1}^{N}\phi_{i}^{*}$, with
\begin{equation}
\begin{array}{l}
\phi _{i}^{\ast }=\sup_{\left( \mbox{\scriptsize\boldmath
$\dot{\varepsilon}$}^{\prime},\mbox{\scriptsize\boldmath
$\dot{\varepsilon}$}
_{i}^{p^{\prime}},\dot{\zeta}_{i}^{\prime}\right) \mathbf{\in
}\bar{\mathbb{D}}_{i}}\left( \mbox{\boldmath
$\sigma$}^{d^{\prime}}:\mbox{\boldmath
$\dot{\varepsilon}$}^{\prime}+\mbox{\boldmath
$\tau$}_{i}^{d^{\prime}}: \mathbf{\mbox{\boldmath
$\dot{\varepsilon}$}}_{i}^{p^{\prime}}+
R_{i}^{d}\dot{\zeta}_{i}^{\prime}-\phi _{i}\right) \\
\ \ \ \ \ = \mathbb{I}_{0}\left( \mbox{\boldmath
$\sigma$}^{d^{\prime}}\right)+\mathbb{I}_{ \mathbb{E}_{i}}\left(
\mbox{\boldmath
$\tau$}^{d^{\prime}}_{i},R^{d^{\prime}}_{i};\mbox{\boldmath
$\varepsilon$},\mbox{\boldmath $\varepsilon$} _{i}^{p},\zeta_{i}
\right)
\end{array}
\end{equation}
where $\mathbb{E}_{i}=\left\{ \left( \mbox{\boldmath
$\tau$}_{i}^{d^{\prime}},R_{i}^{d^{\prime}}\right) \in
\mathbb{S}^{2}\mathbb{\times R}\textrm{ \ \ such \ that \ \
}f_{i}\left( \mbox{\boldmath
$\tau$}_{i}^{d^{\prime}},R_{i}^{d^{\prime}};\mbox{\boldmath
$\varepsilon$},\mbox{\boldmath
$\varepsilon$}^{p}_{i},\zeta_{i}\right)\leq 0\right\} $ and
\begin{equation}
\begin{array}{c}
f_{i}=\frac{dev\left( \mbox{\footnotesize\boldmath
$\tau$}_{i}^{d^{\prime}}\right) :dev\left[ \mathbf{C} :\left(
\mbox{\footnotesize\boldmath$\varepsilon-\varepsilon$}_{i}^{p}\right)
\right] }{2G \ g_{i}\left( \zeta _{i}\right) /\beta
_{i}}-\frac{\left\Vert dev\left[
\mathbf{C}:\left(\mbox{\footnotesize\boldmath
$\varepsilon-\varepsilon$}_{i}^{p}\right) \right] \right\Vert
^{2}}{2G \ g_{i}\left( \zeta _{i}\right) /\beta _{i}}
+R_{i}^{d^{\prime}}
\end{array}
\end{equation}
The flow rules then become of the form (\ref{flowendoIso}).
Moreover, it can be easily proved that at the actual state
represented by $(\mbox{\boldmath $\tau$}_{i}^{d},R_{i}^{d})$, the
identities $f_{i}=\dot{f}_{i}=0$ hold and, for this reason, the
fluxes $\dot{\zeta} _{i}=\dot{\lambda}_{i}\geq 0$ cannot be
computed from the consistency conditions and have to be defined
using a further assumption.

If the number of elements is $N=2$, $ g_{1}=g_{2}=1$ and both
fluxes $\dot{\zeta}_{1}$ and $\dot{\zeta}_{2}$ are of the form
(\ref{BoucWen}), then the model of \citet{Casciati89} is
retrieved. Moreover, the condition
$\dot{\zeta}_{i}^{\prime}=\left\Vert \mathbf{\mbox{\boldmath
$\dot{\varepsilon}$}}_{i}^{p^{\prime}}\right\Vert $ into
(\ref{fimultiendo}) leads to a multi-layer model of Prandtl-Reuss
type.

\section{Non-linear kinematic hardening models}

The NLK hardening rule was first suggested by \citet
{Armstrong66}, who introduced a \emph{dynamic recovery} term in
the classical Prager's linear kinematic hardening rule. Several
modifications of this basic rule have been proposed, in order to
improve the description of the cyclic behavior of metals,
particularly for the ratchetting phenomenon (see, among others,
\cite{Chaboche91} and \cite{Ohno93}).

According to traditional formulation, NLK hardening models
\emph{do not} fulfil the assumption of generalized normality
\cite[pp. 219-221]{Lemaitre90engl} \citep{Chaboche95}. Following
an approach based on the notion of bipotential, \cite{DeSaxce92}
introduced \emph{implicit standard materials} and showed that the
plasticity models with NLK hardening rules are of such type.

\begin{figure}
\begin{center}
\includegraphics[width=13.4cm]{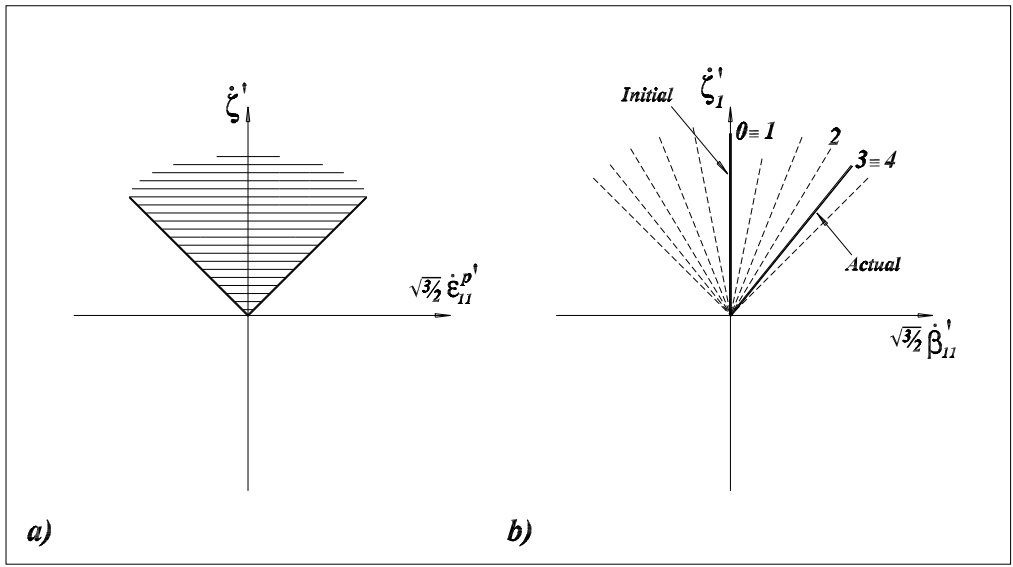}
\caption{NLK hardening model. Tension-compression case. \newline
a) Projection of the effective domain $\bar{\mathbb{D}}$ on the
$\left(\mbox{\boldmath
$\dot{\varepsilon}$}^{p^{\prime}},\dot{\zeta}^{\prime}
\right)$-plane. b) Projection of $\bar{\mathbb{D}}$ on the
$\left(\mbox{\boldmath
$\dot{\beta}$}^{\prime},\dot{\zeta}_{1}^{\prime} \right)$-plane.}

\label{FigNLK1}
\end{center}
\end{figure}

In this section, another formulation is suggested, which leads to
the proof that NLK hardening models belong to the class of
generalized standard materials, provided that a suitable,
non-conventional, loading function is defined. First, the state
variables $\mathbf{v}=\left( \mbox{\boldmath $\varepsilon$}
,\mbox{\boldmath $\varepsilon $}^{p},\zeta ,\mbox{\boldmath $\beta
$},\zeta _{1}\right) $ have to be introduced. The first three are
the same as for Prandtl-Reuss and endochronic models, while
$\mbox{\boldmath $\beta$}$ and $\zeta _{1}$ are related to NLK
hardening rule. The role of the scalar variable $\zeta_{1}$ will
be discussed later on. The corresponding thermodynamic forces are
$ \mathbf{q}^{nd}=\left( \mbox{\boldmath
$\sigma$}^{nd},\mbox{\boldmath $\tau$}^{nd},R^{nd},
\mathbf{X}^{nd},R_{1}^{nd}\right) $ and $\mathbf{q}^{d}=\left(
\mbox{\boldmath $\sigma$}^{d},\mbox{\boldmath
$\tau$}^{d},R^{d},\mathbf{X}^{d},R_{1}^{d}\right) \in
\mathbb{V}^{\ast }$. The Helmholtz energy density is chosen as
follows:
\begin{equation}
\Psi =\frac{1}{2}\left( \mbox{\boldmath
$\varepsilon-\varepsilon$}^{p}\right) : \mathbf{C}:\left(
\mbox{\boldmath $\varepsilon-\varepsilon$}^{p}\right) +\frac{1}{2}
\left( \mbox{\boldmath $\varepsilon$}^{p}-\mbox{\boldmath
$\beta$}\right) :\mathbf{D}:\left( \mbox{\boldmath
$\varepsilon$}^{p}-\mbox{\boldmath $\beta$}\right) \label{NLKhelm}
\end{equation}
The quantity $\mbox{\boldmath $\alpha $}=\mbox{\boldmath
$\varepsilon$}^{p}-\mbox{\boldmath $\beta$}$ is usually adopted as
the internal variable associated with the kinematic hardening.
However, the choice of $\mbox{\boldmath $\beta$}$ as a
representative internal variable appears more suited, because it
highlights the formal analogy between the first quadratic term in
Eq. (\ref{NLKhelm}), typical of plasticity models, and the second
one, associated with the kinematic hardening. The isotropy
assumption leads to the usual expression for $ \mathbf{C}$ and
entails that $\mathbf{D=}D_{1}\mathbf{1\otimes 1+}D_{2}
\mathbf{I}$. The non-dissipative forces can then be readily
evaluated:
\begin{equation}
\begin{array}{l}
\mbox{\boldmath $\sigma$}^{nd}=\mathbf{C}:\left( \mbox{\boldmath
$\varepsilon-\varepsilon$}
^{p}\right) \\
\mbox{\boldmath $\tau$}^{nd}=-\mathbf{C}:\left( \mbox{\boldmath
$\varepsilon-\varepsilon$} ^{p}\right) +\mathbf{D}:\left(
\mbox{\boldmath $\varepsilon$}^{p}-\mbox{\boldmath $\beta$}
\right) , \ \ \ \ \ \ \ \ \ \ \ \ \ \ \ \ R^{nd}=0 \\
\mathbf{X}^{nd}=-\mathbf{D}:\left( \mbox{\boldmath
$\varepsilon$}^{p}-\mbox{\boldmath $\beta$} \right) , \ \ \ \ \ \
\ \ \ \ \ \ \ \ \ \ \ \ \ \ \ \ \ \ \ \ \ \ \ \ \ \ \ \ \
R_{1}^{nd}=0
\end{array}
\label{NLKnondiss}
\end{equation}
The three tensorial non-dissipative forces are related by the
identity $\mbox{\boldmath $\tau$}^{nd}=- \mbox{\boldmath
$\sigma$}^{nd}-\mathbf{X} ^{nd} .$ Moreover, let the
pseudo-potential be equal to
\begin{equation}
\begin{array}{l}
\phi =\sqrt{\frac{2}{3}}\sigma _{y} \ g\left( \zeta \right) \
\dot{\zeta}^{\prime}+\frac{\left\Vert \mathbf{D}:\left(
\mbox{\footnotesize\boldmath$\varepsilon$}^{p}-
\mbox{\footnotesize\boldmath$\beta$}\right) \right\Vert
^{2}}{\frac{D_{2}}{\delta } \ g_{1}\left( \zeta _{1}\right) } \
\dot{\zeta}_{1}^{\prime} \\ \ \ \ \ \ \ \ \ \ \ \ \ \
+\mathbb{I}_{\bar{\mathbb{D}}}\left( \mbox{\boldmath
$\dot{\varepsilon}$}^{\prime},\mbox{\boldmath
$\dot{\varepsilon}$}^{p^{\prime}},\dot{\zeta}^{\prime},\mbox{\boldmath
$\dot{\beta}$}^{\prime},\dot{\zeta} _{1}^{\prime}; \mbox{\boldmath
$\varepsilon$},\mbox{\boldmath
$\varepsilon$}^{p},\zeta,\mbox{\boldmath $\beta$},\zeta
_{1} \right) \\
 \ \  \\
\ \bar{\mathbb{D}}\mathbf{=}\left\{
\begin{array}{l}
\left( \mbox{\boldmath
$\dot{\varepsilon}$}^{\prime},\mbox{\boldmath
$\dot{\varepsilon}$}^{p^{\prime}},\dot{\zeta}^{\prime},\mbox{\boldmath
$\dot{\beta}$}^{\prime},\dot{\zeta}_{1}^{\prime}\right) \in \mathbb{V}\textrm{ \ \ such \ that} \\
\ tr\left( \mbox{\boldmath
$\dot{\varepsilon}$}^{p^{\prime}}\right) =0, \ \ \ \
\dot{\zeta}^{\prime}\geq \left\Vert \mbox{\boldmath
$\dot{\varepsilon}$}^{p^{\prime}}\right\Vert ,
 \ \  \\
\ tr\left( \mbox{\boldmath $\dot{\beta}$}^{\prime}\right) =0, \ \
\ \ \mbox{\boldmath $\dot{\beta}$}^{\prime}=
\frac{\mathbf{D}:\left( \mbox{\footnotesize\boldmath
$\varepsilon$} ^{p}-\mbox{\footnotesize\boldmath $\beta$}\right)
}{\frac{D_{2}}{\delta } \ g_{1}\left( \zeta _{1}\right) } \
\dot{\zeta}_{1}^{\prime}\textrm{ , }
\\ \ \dot{\zeta}_{1}^{\prime}=h\left( \mbox{\boldmath
$\varepsilon$},\mbox{\boldmath $\varepsilon$} ^{p},\zeta
,\mbox{\boldmath $\beta$},\zeta _{1}\right)
\dot{\zeta}^{\prime}\geq 0
\end{array}
\right\}
\end{array}
\label{NLKpseudo}
\end{equation}
with $\delta ,g\left( \zeta \right) ,g_{1}\left( \zeta _{1}\right)
>0$ and $g\left(0 \right)=g_{1}\left(0 \right)=1$. Figure
\ref{FigNLK1} shows two projections of the effective domain
$\bar{\mathbb{D}}$ for the tension-compression case. The first
term in the definition of $\phi $ is identical to that of Eq.
(\ref{PseudoReussIso1}) for a modified Prandtl-Reuss model with
$\xi(\zeta)=0$. The second term is related to the NLK hardening
and \emph{it is formally identical to the one used in the
definition of endochronic models} (see Eq. (\ref{fiEndoIso})),
with the substitutions $ dev\left(\mbox{\boldmath
$\varepsilon$}\right)\rightarrow \mbox{\boldmath
$\varepsilon$}^{p}$, $\mbox{\boldmath
$\varepsilon$}^{p}\rightarrow \mbox{\boldmath $\beta$}$ and $\zeta
\rightarrow \zeta_{1}$. The same analogy applies to the conditions
defining the set $\bar{\mathbb{D}}$.

\begin{figure}
\begin{center}
\includegraphics[width=13.4cm]{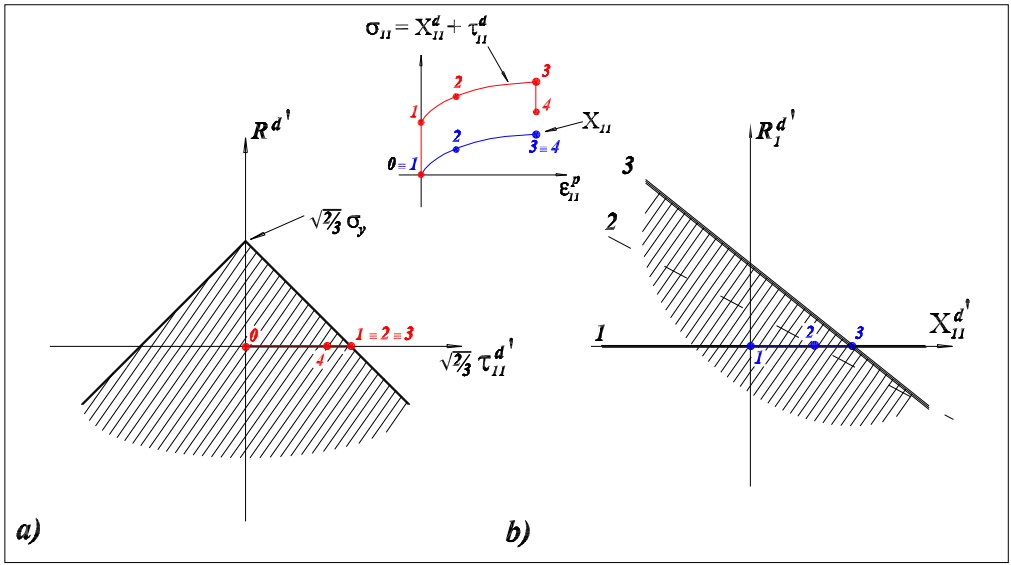}
\caption{NLK hardening model. Tension-compression case with
$g(\zeta)=1$. \newline a) Condition $f\leq 0$ when
$R^{d^{\prime}}_{1}=R^{d}_{1}=0$ and
$\mathbf{X}^{d^{\prime}}=\mathbf{X}^{d}=\mathbf{D:}\left(\mbox{\boldmath
$\varepsilon$}^{p}-\mbox{\boldmath $\beta$}\right)$. b) Condition
$f\leq 0$ when $R^{d^{\prime}}=R^{d}=0$ and
$\|dev\left(\mbox{\boldmath
$\tau$}^{d^{\prime}}\right)\|=\|dev\left(\mbox{\boldmath
$\tau$}^{d}\right)\|=\sqrt{\frac{2}{3}}\sigma_{y} $.}

\label{FigNLK2}
\end{center}
\end{figure}

The dual pseudo-potential then becomes
\begin{equation}
\begin{array}{l}
\phi ^{\ast }=\sup_{\left(\mbox{\scriptsize\boldmath
$\dot{\varepsilon}$}^{\prime},\mbox{\scriptsize\boldmath
$\dot{\varepsilon}$}^{p^{\prime}},\dot{\zeta}^{\prime},\mbox{\scriptsize\boldmath
$\dot{\beta}$}^{\prime},\dot{\zeta}_{1}^{\prime}\right) \in
\bar{\mathbb{D}}}\left(
\begin{array}{l}
\mbox{\boldmath $\sigma$}^{d^{\prime}}:\mbox{\boldmath$
\dot{\varepsilon}$}^{\prime}+\mbox{\boldmath$\tau$}^{d^{\prime}}:\mbox{\boldmath
$\dot{\varepsilon}$}^{p^{\prime}}+R^{d^{\prime}}\dot{\zeta}^{\prime}+
\\ \ \ \ \ \ \ \ \ \ \  \ \  \ \ \ \mathbf{X}^{d^{\prime}}:\mbox{\boldmath$\dot{\beta}$}^{\prime}
+R_{1}^{d^{\prime}}\dot{\zeta}_{1}^{\prime}-\phi
\end{array}
\right)  \\
=\mathbb{I}_{0}\left( \mbox{\boldmath $\sigma$}^{d^{\prime}}
\right)+\mathbb{I}_{\mathbb{E}}\left( \mbox{\boldmath
$\tau$}^{d^{\prime}},R^{d^{\prime}},\mathbf{X}^{d^{\prime}},R_{1}^{d^{\prime}};\mbox{\boldmath
$\varepsilon$}, \mbox{\boldmath
$\varepsilon$}^{p},\zeta,\mbox{\boldmath $\beta$},\zeta_{1}
\right)
\end{array}
\label{NLKfistar}
\end{equation}
where $\mathbb{E}=\left\{ \left( \mbox{\boldmath
$\tau$}^{d^{\prime}},R^{d^{\prime}},\mathbf{X}
^{d^{\prime}},R_{1}^{d^{\prime}}\right) \in
\mathbb{S}^{2}\mathbb{\times R\times S}^{2}\mathbb{ \times
R}\textrm{ \ \ such \ that \ \ }f\leq 0\right\}$  and
\begin{equation}
\begin{array}{l}
f=\left\Vert dev\left( \mbox{\boldmath $\tau$}^{d^{\prime}}\right)
\right\Vert -\sqrt{\frac{2}{
3}}\sigma _{y} \ g\left( \zeta \right) +R^{d^{\prime}} \\
 \ \ \ \ \ \ \ \ \ \ \ \ +\left( \frac{\mathbf{X}^{d^{\prime}}:\left[ \mathbf{D
}:\left( \mbox{\footnotesize\boldmath
$\varepsilon$}^{p}-\mbox{\footnotesize\boldmath $\beta$}\right) \
\right] }{ D_{2} \ g_{1}\left( \zeta _{1}\right) /\delta
}-\frac{\left\Vert \mathbf{D} :\left( \mbox{\footnotesize\boldmath
$\varepsilon$}^{p}-\mbox{\footnotesize\boldmath $\beta$}\right)
\right\Vert ^{2}}{D_{2} \ g_{1}\left( \zeta _{1}\right) /\delta
}+R_{1}^{d^{\prime}}\right) h\left(\mbox{\boldmath
$\varepsilon$},\mbox{\boldmath $\varepsilon$}^{p},
\zeta,\mbox{\boldmath $\beta$},\zeta_{1} \right)
\end{array}
\label{NLKloadf}
\end{equation}
Eq. (\ref{NLKloadf}) defines the loading function of a model with
NLK hardening and the associated set $\mathbb{E}$ is depicted in
Figure \ref{FigNLK2} for the tension-compression case when
$g(\zeta)=1$. The normality condition associated with
$\phi^{\ast}$ leads to the following flow rules:
\begin{equation}
\begin{array}{l}
\begin{array}{l}
\mbox{\boldmath $\dot{\varepsilon}$}^{p}=\frac{dev\left(
\mbox{\footnotesize\boldmath $\tau$}^{d}\right) }{ \left\Vert
dev\left( \mbox{\footnotesize\boldmath $\tau$}^{d}\right)
\right\Vert }\dot{\lambda}= \mathbf{n}\dot{\lambda} \ \ \ \ \ \ \
\ \ \ \ \ \ \ \ \ \ \ \ \ \ \ \dot{\zeta}=\dot{\lambda
} \\
\mbox{\boldmath $\dot{\beta}$}=\frac{\mathbf{D}:\left(
\mbox{\footnotesize\boldmath $\varepsilon$}^{p}-
\mbox{\footnotesize\boldmath $\beta$}\right) }{D_{2} \ g_{1}\left(
\zeta _{1}\right) /\delta } \ h\left(\mbox{\boldmath
$\varepsilon$},\mbox{\boldmath $\varepsilon$}^{p},
\zeta,\mbox{\boldmath $\beta$},\zeta_{1} \right) \ \dot{\lambda} \
\ \ \ \ \ \ \ \dot{\zeta} _{1}=h\left(\mbox{\boldmath
$\varepsilon$},\mbox{\boldmath $\varepsilon$}^{p},
\zeta,\mbox{\boldmath $\beta$},\zeta_{1} \right) \dot{\lambda}
\end{array}
\\
\textrm{with \ \ \ } \dot{\lambda}\geq 0, \ \ \ f\leq 0, \ \ \dot{
\lambda}f=0
\end{array}
\label{NLKflow}
\end{equation}
The thermodynamic force
$\mathbf{X}^{d}=-\mathbf{X}^{nd}=\mathbf{D}:\left( \mbox{\boldmath
$\varepsilon$}^{p}- \mbox{\boldmath $\beta$}\right) $ is
traceless, due to the assumptions adopted for the traces of
$\mbox{\boldmath $\dot{\varepsilon}$}^{p}$ and $\mbox{\boldmath
$\dot{\beta}$}$. Special attention must be paid to the
relationship between the fluxes $\dot{\zeta}_{1}$ and
$\dot{\zeta}$. The time derivative of $\zeta_{1}$ is defined as
the product between $\dot{\zeta}$ and the function $h$, which
depends on the state variables and must be non-negative and
finite, but is otherwise free. The variable $\zeta_{1}$ can be
interpreted as an \emph{intrinsic time scale for the NLK hardening
flow rule}.

Accounting for the identities $\left( \mbox{\boldmath $\tau$}
^{nd},R^{nd},\mathbf{X}^{nd},R_{1}^{nd}\right) =-\left(
\mbox{\boldmath $\tau$} ^{d},R^{d},\mathbf{X}^{d},R_{1}^{d}\right)
$ and Eqs. (\ref{NLKnondiss}), one can prove that $R^{d}=0$ and
that the term proportional to $h$ in Eq. (\ref{NLKloadf}) is
always zero at the actual state. Hence, only the first two terms
in the expression of $f$ affect the consistency condition
$\dot{f}=0$, which leads to the plastic multiplier
\begin{equation}
\dot{\lambda}=H\left( f\right) \frac{\left\langle
\mathbf{n}:\mathbf{\mbox{\boldmath
$\dot{\varepsilon}$}}\right\rangle
}{1\mathbf{+}\frac{D_{2}}{2G}-\frac{1}{2G}
\frac{\mathbf{n:X}^{d}}{g_{1}\left( \zeta _{1}\right) /\delta }
h\left(\mbox{\boldmath $\varepsilon$},\mbox{\boldmath
$\varepsilon$}^{p}, \zeta,\mbox{\boldmath $\beta$},\zeta_{1}
\right) \ +\sqrt{\frac{2}{3}}\frac{\sigma _{y}}{2G} \
\frac{dg\left( \zeta \right)}{d\zeta} }
\end{equation}
The positive functions $g,g_{1}$ and $h$ determine the actual
model.

The choice $g=g_{1}=h=1$ corresponds to the basic NLK hardening
model of \citet{Armstrong66}. Another interesting case is given by
$g=g_{1}=1$ and
\begin{equation}
\left\{
\begin{array}{ll}
h=\left( \frac{\left\Vert \mathbf{D}:\left(
\footnotesize{\mbox{\boldmath
$\varepsilon$}^{p}}-\footnotesize{\mbox{\boldmath $\beta$}}\right)
\right\Vert }{D_{2}/\delta }\right) ^{m_{1}}\left\langle
\mathbf{k}_{1}:\mathbf{n}\right\rangle  & \textrm{ \ \ if \ \
}\mathbf{D}:\left( \mbox{\boldmath
$\varepsilon$}^{p}-\mbox{\boldmath $\beta$}\right) \neq
\mathbf{0} \\
h=0 & \textrm{ \ \ if \ \ }\mathbf{D}:\left( \mbox{\boldmath
$\varepsilon$}^{p}-\mbox{\boldmath $\beta$}\right) = \mathbf{0}
\end{array}
\right. \label{hOhno}
\end{equation}
where $m_{1}>0$ and $\mathbf{k}_{1}=\frac{\mathbf{D}:\left(
\mbox{\footnotesize\boldmath
$\varepsilon$}^{p}-\mbox{\footnotesize\boldmath $\beta$}\right)
}{\left\Vert \mathbf{D}:\left( \mbox{\footnotesize\boldmath
$\varepsilon$}^{p}-\mbox{\footnotesize\boldmath $\beta$}\right)
\right\Vert }$ is the unit vector having the same direction as
$\mathbf{X}^{d}=\mathbf{D}:\left( \mbox{\boldmath
$\varepsilon$}^{p}-\mbox{\boldmath $\beta$}\right)$. These
conditions lead to
\begin{equation}
\mbox{\boldmath $\dot{\beta}$}=\frac{\mathbf{X}^{d}}{D_{2}/\delta
}\dot{\zeta}_{1}=\frac{ \mathbf{X}^{d}}{D_{2}/\delta }\left(
\frac{\left\Vert \mathbf{X} ^{d}\right\Vert }{D_{2}/\delta
}\right) ^{m_{1}}\left\langle \mathbf{k}_{1}: \mbox{\boldmath
$\dot{\varepsilon}$}^{p}\right\rangle =\mbox{\boldmath
$\dot{\varepsilon}$}^{p}- \frac{\mathbf{\dot{X}}^{d}}{D_{2}}
\label{betaOhno}
\end{equation}
which is the NLK hardening rule proposed by \cite {Ohno93} for
modelling the ratchetting phenomenon in metal plasticity. It is
interesting to compare the quantity
\begin{equation}
\dot{\zeta}_{1}=h \ \dot{\zeta}=\left( \frac{\left\Vert \mathbf{X}
^{d}\right\Vert }{D_{2}/\delta }\right) ^{m_{1}}\left\langle
\mathbf{k}_{1}: \mbox{\boldmath
$\dot{\varepsilon}$}^{p}\right\rangle
\end{equation}
and the intrinsic time flow $\dot{\vartheta}$, defined in Eq.
(\ref {BoucWen}) for endochronic models of the Bouc-Wen type. Two
significant differences can be observed: (i) the governing flow
variable is the plastic strain for NLK hardening rule and the
total strain for the flow rule of the endochronic model; (ii) due
to presence of the \emph{absolute value} instead of the McCauley
brackets, the endochronic model of Bouc-Wen type introduces
non-zero flows during unloading phases when $\gamma \neq \beta$.

\subsection{From an endochronic model to a NLK hardening model}

\cite{Valanis80} and \cite{Watanabe86} proved that a NLK hardening
model can be derived from the endochronic theory by adopting a
special intrinsic-time definition, namely when the intrinsic time
scale flow $\dot{\zeta}$ is forced to be equal to the norm of the
plastic strain flow. The approach suggested in this paper not only
confirms this result, but allows for a generalization, due to the
presence of a second intrinsic time scale $\zeta _{1}$, in general
distinct from $\zeta$. Consider the differential equations
defining an endochronic model with a $\emph{kinematic hardening}$
variable $\mbox{\boldmath $\tau$}^{d}:$
\begin{equation}
\left\{
\begin{array}{l}
tr\left( \mbox{\boldmath $\dot{\sigma}$}\right)
 =3K \ tr\left( \mathbf{\mbox{\boldmath $\dot{\varepsilon}$}}\right) \\
dev\left( \mbox{\boldmath $\dot{\sigma}$}\right) =2G \
dev(\mbox{\boldmath $\dot{\varepsilon}$})-\beta \ dev\left(
\mbox{\boldmath $\sigma -\tau $}^{d}\right) \frac{\dot{\zeta}}{
g\left( \zeta \right) } \\
tr\left( \mbox{\boldmath $\dot{\tau}$}^{d}\right) =0 \\
\mbox{\boldmath $\dot{\tau}$}^{d}=D_{2}\mbox{\boldmath
$\dot{\varepsilon}$}^{p}\mathbf{-}\delta \ \mbox{\boldmath
$\tau$}^{d}\frac{\dot{\zeta}_{1}}{g_{1}\left( \zeta
_{1}\right) } \\
\dot{\zeta}_{1}=h\left( \mbox{\boldmath
$\varepsilon$},\mbox{\boldmath $\varepsilon$}^{p},\zeta ,
\mbox{\boldmath $\beta$},\zeta _{1}\right) \ \dot{\zeta}
\end{array}
\right.  \label{endoNLK}
\end{equation}
The idea of a kinematic hardening variable in an endochronic model
was first suggested by \citet{Bazant78}, who however considered a
\emph{linear} evolution of $\mbox{\boldmath $\tau$}^{d}$ as
function of the plastic strain. An alternative way to describe the
model defined by (\ref{endoNLK}) is
\begin{equation}
\left\{
\begin{array}{ll}
\mbox{\boldmath $\sigma$}=\mathbf{C}:\left( \mbox{\boldmath
$\varepsilon$}-\mbox{\boldmath $\varepsilon$}^{p}\right)  &
\mbox{\boldmath $\tau$}^{d}=\mathbf{D}:\left( \mbox{\boldmath
$\varepsilon$}
^{p}-\mbox{\boldmath $\beta$}\right)  \\
\mathbf{C=}\left( K-\frac{2}{3}G\right) \mathbf{1\otimes
1}+2G\mathbf{I}
\textrm{ \ \ } & \mathbf{D=}D_{1}\mathbf{1\otimes 1}+D_{2}\mathbf{I} \\
tr\left( \mbox{\boldmath $\dot{\varepsilon}$}^{p}\right)
=0,\textrm{ \ \ }\mbox{\boldmath
$\dot{\varepsilon}$}^{p}=\frac{dev(\mbox{\footnotesize\boldmath
$\sigma-\tau $}^{d})}{\frac{2G}{\beta }g\left( \zeta \right)
}\dot{\zeta}\textrm{ \ \ \ \ } & tr\left( \mbox{\boldmath
$\dot{\beta}$}\right)=0,\textrm{ \ \ }\mbox{\boldmath
$\dot{\beta}$}=\frac{\mbox{\footnotesize\boldmath
$\tau$}^{d}}{\frac{D_{2}}{
\delta }g_{1}\left( \zeta _{1}\right) }\dot{\zeta}_{1} \\
& \dot{\zeta}_{1}=h\left( \mbox{\boldmath
$\varepsilon$},\mbox{\boldmath $\varepsilon$}^{p}, \zeta,
\mbox{\boldmath $\beta$},\zeta _{1}\right)\dot{\zeta}
\end{array}
\right.
 \label{endoNLK1}
\end{equation}

Moreover, both Eqs. (\ref{endoNLK}) and (\ref{endoNLK1}) can be
derived from (\ref{NLKhelm}) and the following pseudo-potential:
\begin{equation}
\begin{array}{l}
\phi =\frac{\left\Vert dev\left[
\mathbf{C}:\left(\mbox{\footnotesize\boldmath
$\varepsilon-\varepsilon$}^{p}\right)-\mathbf{D}:\left(
\mbox{\footnotesize\boldmath$\varepsilon$}
^{p}-\mbox{\footnotesize\boldmath$\beta$}\right)\right]
\right\Vert ^{2}}{\frac{2G}{\beta} \ g\left( \zeta \right) } \
\dot{\zeta}^{\prime}+\frac{\left\Vert \mathbf{D}
:\left(\mbox{\footnotesize\boldmath
$\varepsilon$}^{p}-\mbox{\footnotesize\boldmath $\beta$}\right)
\right\Vert ^{2}}{ \frac{D_{2}}{\delta } \ g_{1}\left( \zeta
_{1}\right) } \ \dot{
\zeta}_{1}^{\prime} \\
\ \ \ \ \ \ +\mathbb{I}_{\bar{\mathbb{D}}}\left( \mbox{\boldmath
$\dot{\varepsilon }$}^{\prime},\mbox{\boldmath
$\dot{\varepsilon}$}^{p^{\prime}},\dot{\zeta}^{\prime},\mbox{\boldmath
$\dot{\beta}$}^{\prime},\dot{\zeta} _{1}^{\prime};\mbox{\boldmath
$\varepsilon$},\mbox{\boldmath
$\varepsilon$}^{p},\zeta,\mbox{\boldmath $\beta$},\zeta
_{1}\right) \\
 \ \  \\
\ \bar{\mathbb{D}}\mathbf{=}\left\{
\begin{array}{l}
\left( \mbox{\boldmath
$\dot{\varepsilon}$}^{\prime},\mbox{\boldmath
$\dot{\varepsilon}$}^{p^{\prime}},\dot{\zeta}^{\prime},\mbox{\boldmath
$\dot{\beta}$}^{\prime},\dot{\zeta}_{1}^{\prime}\right) \in \mathbb{V}\textrm{ \ \ \ such \ that} \\
 tr\left( \mbox{\boldmath $\dot{\varepsilon}$}^{p^{\prime}}\right) =0,
 \ \ \ \mbox{\boldmath $\dot{\varepsilon}$}^{p^{\prime}}\mathbf{=} \ \frac{dev
\left[ \mathbf{C}:\left( \mbox{\footnotesize\boldmath
$\varepsilon-\varepsilon$}^{p}\right) -\mathbf{D}:\left(
\mbox{\footnotesize\boldmath
$\varepsilon$}^{p}-\mbox{\footnotesize\boldmath $\beta$}\right)
\right] }{\frac{2G}{\beta } \ g\left( \zeta \right) } \
\dot{\zeta}^{\prime},
 \ \ \ \dot{\zeta}^{\prime}\geq 0, \\ tr\left( \mbox{\boldmath
$\dot{\beta}$}^{\prime} \right) =0\textrm{,} \ \mbox{\boldmath
$\dot{\beta}$}^{\prime}=
\frac{\mathbf{D}:\left(\mbox{\footnotesize\boldmath$\varepsilon$
}^{p}-\mbox{\footnotesize\boldmath $\beta$}\right)
}{\frac{D_{2}}{\delta } \ g_{1}\left( \zeta _{1}\right) } \
\dot{\zeta}_{1}^{\prime}\textrm{, \ }
\dot{\zeta}_{1}^{\prime}=h\left( \mbox{\boldmath
$\varepsilon$},\mbox{\boldmath $\varepsilon$}^{p},\zeta ,
\mbox{\boldmath $\beta$},\zeta _{1}\right)
\dot{\zeta}^{\prime}\geq 0
\end{array}
\right\}
\end{array}
\label{fiendoNLK}
\end{equation}
Let $\dot{\zeta}^{\prime}=\left\Vert \mbox{\boldmath
$\dot{\varepsilon}$}^{p^{\prime}}\right\Vert $ be the chosen
intrinsic time definition and assume
$\frac{2G}{\beta}=\sqrt{\frac{2}{3}}\sigma_{y}$. Then, introducing
these conditions in (\ref{fiendoNLK}), one obtains a
pseudo-potential $\tilde{\phi}$ which differs from the one of Eq.
(\ref {NLKpseudo}) only in the inequality
$\dot{\zeta}^{\prime}\geq \left\Vert \mathbf{\mbox{\boldmath
$\dot{\varepsilon}$}}^{p^{\prime}}\right\Vert $, which is an
equality in $\tilde{\phi}$. This difference affects neither the
expression of the dual pseudo-potential $\tilde{\phi ^{\ast
}}=\phi ^{\ast }$ (Appendix, item 6) nor the flow rules, which are
in both cases equal to Eqs. (\ref{NLKfistar})-(\ref{NLKloadf}) and
Eq. (\ref {NLKflow}), respectively. Moreover, in the particular
case $h=1$\ and $g\left( \zeta \right) =g_{1}\left( \zeta \right)
$, the results discussed by \cite{Valanis80} and \cite{Watanabe86}
are retrieved.

\section{Generalized plasticity models}

Generalized plasticity models \citep {Lubliner93} are considered
an effective alternative to NLK hardening models, since they
behave similarly and are computationally less expensive
\citep{Auricchio95}. A new description of these models is
suggested here, supported by a suitable pseudo-potential and the
generalized normality assumption. In order to expose the basic
principles of this new approach, only the simple generalized
plasticity model presented by \citet{Auricchio95} is considered.
The extension to more complex cases is straightforward.

\begin{figure}
\begin{center}
\includegraphics[width=13.4cm]{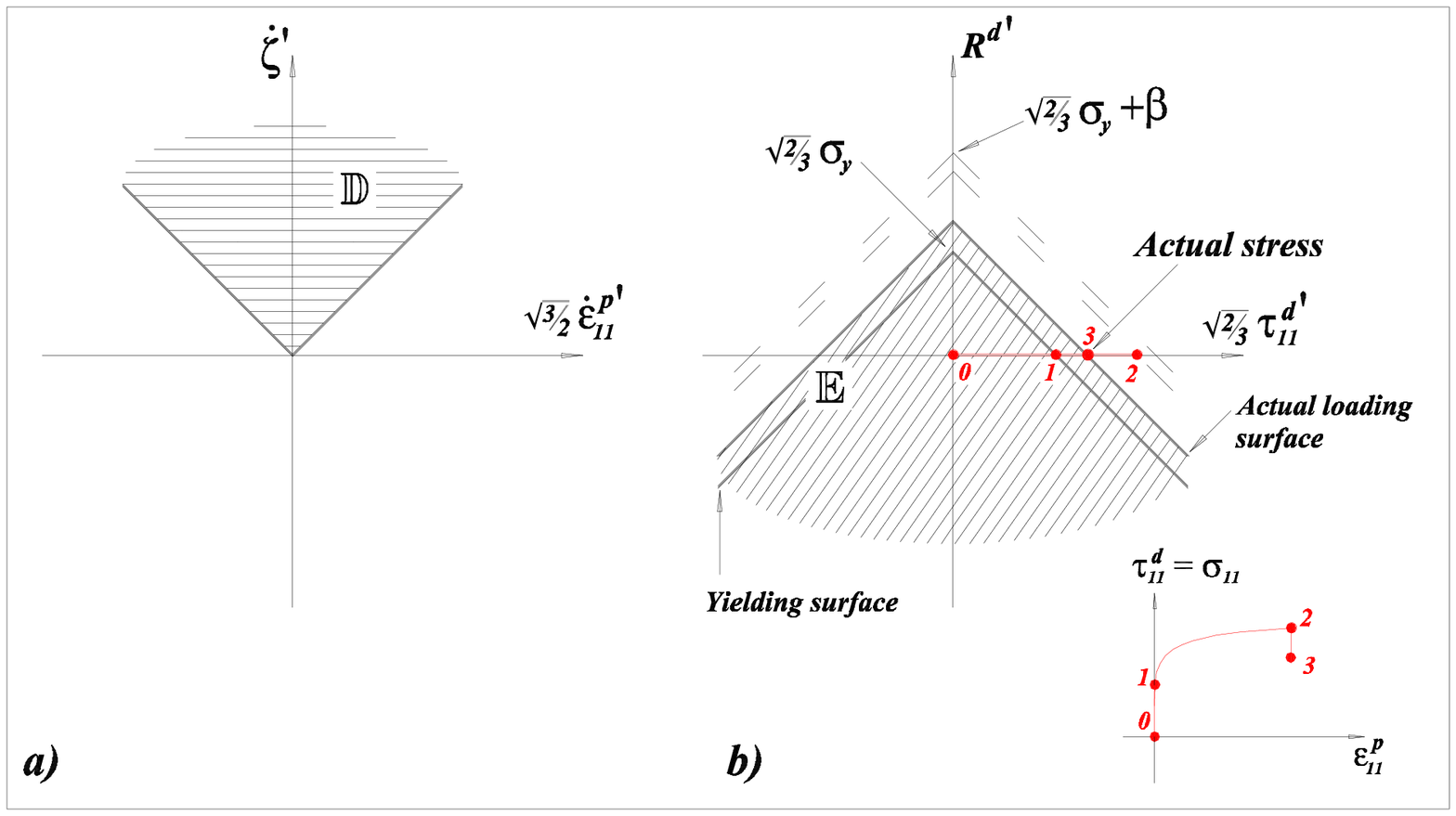}
\caption{Generalized plasticity. Tension-compression case.
\newline a) Projection of the
pseudo-potential effective domain $\bar{\mathbb{D}}$ on the
$\left(\mbox{\boldmath
$\dot{\varepsilon}$}^{p^{\prime}},\dot{\zeta}^{\prime}\right)$-plane.
This set is indicated by $\mathbb{D}$. b) Several configurations
of the domain $\mathbb{E}$. When $\bar{f}\geq0$, $\mathbb{E}$
translates upward during loading phases and downward during
unloading phases. The point $(\mbox{\boldmath $\tau$}^{d},R^{d})$,
representing the actual state, always lies on the axis
$R^{d^{\prime}}=0$.}

\label{FigGenPlast}
\end{center}
\end{figure}

First, the state variables $\mathbf{v}=\left( \mbox{\boldmath
$\varepsilon$},\mbox{\boldmath $\varepsilon$}^{p},\zeta \right) $
have to be introduced. The corresponding thermodynamic forces are
$\mathbf{q}^{nd}=\left( \mbox{\boldmath
$\sigma$}^{nd},\mbox{\boldmath $\tau$} ^{nd},R^{nd}\right) $ and
$\mathbf{q}^{d}=\left( \mbox{\boldmath $\sigma$}^{d},
\mbox{\boldmath $\tau$}^{d},R^{d}\right)$. The Helmholtz energy
density is chosen as follows:
\begin{equation}
\Psi =\frac{1}{2}\left( \mbox{\boldmath
$\varepsilon-\varepsilon$}^{p}\right) : \mathbf{C}:\left(
\mbox{\boldmath $\varepsilon-\varepsilon$}^{p}\right) +\frac{1}{2}
\mbox{\boldmath $\varepsilon$}^{p}:\mathbf{D}:\mbox{\boldmath
$\varepsilon$}^{p}
\end{equation}
The expression for $\mathbf{C}$ and $\mathbf{D}$ are the same as
in NLK hardening models. The non-dissipative forces can be readily
evaluated:
\begin{equation}
\begin{array}{c}
\mbox{\boldmath $\sigma$}^{nd}=\mathbf{C}:\left( \mbox{\boldmath
$\varepsilon-\varepsilon$} ^{p}\right) , \ \ \ \mbox{\boldmath
$\tau$}^{nd}=-\mathbf{C}:\left( \mbox{\boldmath
$\varepsilon-\varepsilon$}^{p}\right) +\mathbf{D}:\mbox{\boldmath
$\varepsilon$}^{p}, \ \ \ R^{nd}=0
\end{array}
\end{equation}
Note that $\mbox{\boldmath $\sigma$}^{nd}$ and $\mbox{\boldmath
$\tau$}^{nd}$ are related by the identity $\mbox{\boldmath
$\tau$}^{nd}=-\left( \mbox{\boldmath $\sigma$}^{nd}-\mathbf{D}:
\mbox{\boldmath $\varepsilon$}^{p}\right) ,$ where the
\emph{backstress} $\mathbf{D}:\mbox{\boldmath $\varepsilon$}^{p}$
introduces a linear kinematic hardening effect. Moreover, let the
pseudo-potential be equal to
\begin{equation}
\begin{array}{l}
\phi\left(\mbox{\boldmath
$\dot{\varepsilon}$}^{\prime},\mbox{\boldmath $
\dot{\varepsilon}$}^{p^{\prime}},\dot{\zeta}^{\prime};\mbox{\boldmath
$\varepsilon$},\mbox{\boldmath $\varepsilon$}^{p},\zeta \right) =
\bar{g}\left(\mbox{\boldmath $\varepsilon$},\mbox{\boldmath
$\varepsilon$}^{p},\zeta \right) \
\dot{\zeta}^{\prime}+\mathbb{I}_{\bar{\mathbb{D}}}\left(
\mbox{\boldmath $\dot{\varepsilon}$}^{\prime},\mbox{\boldmath $
\dot{\varepsilon}$}^{p^{\prime}},\dot{\zeta}^{\prime}\right)  \\
\ \  \\
\bar{\mathbb{D}}\mathbf{=}\left\{
\begin{array}{l}
\left( \mbox{\boldmath
$\dot{\varepsilon}$}^{\prime},\mbox{\boldmath
$\dot{\varepsilon}$}^{p^{\prime}},\dot{\zeta}^{\prime}\right) \in
\mathbb{V}\textrm{ \ \ such \ that}  \ \ tr\left( \mbox{\boldmath
$\dot{\varepsilon}$}^{p^{\prime}}\right) =0\textrm{ \ \ and \ \
}\dot{\zeta}^{\prime}\geq \left\Vert \mbox{\boldmath
$\dot{\varepsilon}$}^{p^{\prime}}\right\Vert
\end{array}
\right\}
\end{array}
\label{PseudoGenPl}
\end{equation}
where
\begin{equation}
\begin{array}{l}
\bar{g}\left( \mbox{\boldmath $\varepsilon$},\mbox{\boldmath
$\varepsilon$}^{p},\zeta \right) =\left\{
\begin{array}{ll}
\sqrt{\frac{2}{3}}\sigma _{y}+H_{iso} \ \zeta  & \textrm{ \ \ \ if
\ \ }\bar{f}<0
\\
\left\Vert dev\left[ \mathbf{C}:\left( \mbox{\boldmath
$\varepsilon-\varepsilon$} ^{p}\right) -\mathbf{D}:\mbox{\boldmath
$\varepsilon$}^{p}\right] \right\Vert  & \textrm{ \ \ \ if \ \
}\bar{f}\geq 0
\end{array}
\right.  \\
\bar{f}\left( \mbox{\boldmath $\varepsilon$},\mbox{\boldmath
$\varepsilon$}^{p},\zeta \right):=\left\Vert dev\left[
\mathbf{C}:\left( \mbox{\boldmath
$\varepsilon-\varepsilon$}^{p}\right) -\mathbf{D}:\mbox{\boldmath
$\varepsilon$}^{p}\right] \right\Vert -\left(
\sqrt{\frac{2}{3}}\sigma _{y}+H_{iso} \ \zeta \right)
\end{array}
\end{equation}
with $H_{iso}\geq 0$. The main characteristic of this
pseudo-potential function is given by the \emph{piecewise}
expression introduced to define the positive function $\bar{g}$.
It is assumed that $\bar{g}$ depends on the sign of the function
$\bar{f}$, which in turn is related to the state variables. The
conjugated pseudo-potential $\phi ^{\ast }$ reads
\begin{equation}
\begin{array}{l}
\phi ^{\ast}\left(\mbox{\boldmath
$\sigma$}^{d^{\prime}},\mbox{\boldmath
$\tau$}^{d^{\prime}},R^{d^{\prime}}; \mbox{\boldmath
$\varepsilon$},\mbox{\boldmath $\varepsilon$}^{p},\zeta \right) \\
\ \ \ \ =\sup_{\left( \mbox{\scriptsize\boldmath
$\dot{\varepsilon}$}^{\prime},\mbox{\scriptsize\boldmath
$\dot{\varepsilon}$}^{p^{\prime}},\dot{\zeta}^{\prime}\right)
\mathbf{\in }\bar{\mathbb{D}}}\left( \mbox{\boldmath
$\sigma$}^{d^{\prime}}:\mbox{\boldmath
$\dot{\varepsilon}$}^{\prime}+\mbox{\boldmath
$\tau$}^{d^{\prime}}:\mathbf{\mbox{\boldmath
$\dot{\varepsilon}$}}^{p^{\prime}}+R^{d^{\prime}}\dot{\zeta}^{\prime}-\phi
\right)
\\ \ \ \ \ = \mathbb{I}_{0}\left( \mbox{\boldmath
$\sigma$}^{d^{\prime}}\right)+\mathbb{I}_{\mathbb{E}} \left(
\mbox{\boldmath $\tau$}^{d^{\prime}},R^{d^{\prime}};
\mbox{\boldmath $\varepsilon$},\mbox{\boldmath
$\varepsilon$}^{p},\zeta \right)
\end{array}
\label{fistarGenPl}
\end{equation}
where $\mathbb{E}=\left\{ \left( \mbox{\boldmath
$\tau$}^{d^{\prime}},R^{d^{\prime}}\right) \in \mathbb{S}^{2}
\mathbb{\times R}\textrm{ \ \ such \ that \ \ }f\left(
\mbox{\boldmath $\tau$}^{d^{\prime}},R^{d^{\prime}};
\mbox{\boldmath $\varepsilon$},\mbox{\boldmath
$\varepsilon$}^{p},\zeta \right)\leq 0\right\}$ and
\begin{equation}
\begin{array}{l}
f\left( \mbox{\boldmath $\tau$}^{d^{\prime}},R^{d^{\prime}};
\mbox{\boldmath $\varepsilon$},\mbox{\boldmath
$\varepsilon$}^{p},\zeta \right) \\
\ \ =\left\{
\begin{array}{ll}
\left\Vert dev\left( \mbox{\boldmath $\tau$}^{d^{\prime}}\right)
\right\Vert -\left( \sqrt{ \frac{2}{3}}\sigma _{y}+H_{iso} \ \zeta
\right) +R^{d^{\prime}} & \textrm{ \ if \ }
\bar{f}<0 \\
\left\Vert dev\left( \mbox{\boldmath $\tau$}^{d^{\prime}}\right)
\right\Vert -\left\Vert dev \left[ \mathbf{C}:\left(
\mbox{\boldmath $\varepsilon-\varepsilon$}^{p}\right) -
\mathbf{D}:\mbox{\boldmath $\varepsilon$}^{p}\right] \right\Vert
+R^{d^{\prime}} & \textrm{ \ if \ } \bar{f}\geq 0
\end{array}
\right.
\end{array}
\label{loadFuncGenPl}
\end{equation}
The loading function $f$ also has a twofold definition: recalling
that the actual thermodynamic force $\mbox{\boldmath $\tau$}^{d}$
fulfils the following identities
\begin{equation}
\mbox{\boldmath $\tau$}^{d}=\mathbf{C}:\left( \mbox{\boldmath
$\varepsilon-\varepsilon$} ^{p}\right) -\mathbf{D}:\mbox{\boldmath
$\varepsilon$}^{p}=\mbox{\boldmath
$\sigma$}-\mathbf{D}:\mbox{\boldmath $\varepsilon$}^{p}
\end{equation}
and $R^{d}=-R^{nd}=0$, one can prove that if $\bar{f}\left(
\mbox{\footnotesize\boldmath
$\varepsilon$},\mbox{\footnotesize\boldmath
$\varepsilon$}^{p},\zeta\right)<0$ then $f\left( \mbox{\boldmath
$\tau$}^{d},R^{d}; \mbox{\boldmath $\varepsilon$},\mbox{\boldmath
$\varepsilon$}^{p},\zeta \right)=\bar{f}\left(
\mbox{\footnotesize\boldmath
$\varepsilon$},\mbox{\footnotesize\boldmath
$\varepsilon$}^{p},\zeta\right)$; moreover, if $\bar{ f}\left(
\mbox{\footnotesize\boldmath
$\varepsilon$},\mbox{\footnotesize\boldmath
$\varepsilon$}^{p},\zeta\right)\geq 0$, then $f\left(
\mbox{\boldmath $\tau$}^{d},R^{d}; \mbox{\boldmath
$\varepsilon$},\mbox{\boldmath $\varepsilon$}^{p},\zeta \right)$
is always zero, viz. the actual state represented by
$(\mbox{\boldmath $\tau$}^{d},R^{d})$ remains in contact with the
loading surface $\partial \mathbb{E}$. In Figure
\ref{FigGenPlast}, this situation is depicted for the
tension-compression case.

The normality conditions associated with the loading function $f$
read:
\begin{equation}
\begin{array}{l}
\mbox{\boldmath $\dot{\varepsilon}$}^{p}=\frac{dev\left(
\mbox{\footnotesize\boldmath $\tau$}^{d}\right) }{ \left\Vert
dev\left( \mbox{\footnotesize\boldmath $\tau$}^{d}\right)
\right\Vert }\dot{\lambda}= \mathbf{n}\dot{\lambda}, \ \ \ \ \ \ \
\ \ \dot{\zeta}=\dot{\lambda}
\\
\textrm{with \ \ \ \ \ } \dot{\lambda}f=0 \ \ \ \ \ \ f\leq 0 \ \
\ \ \ \ \dot{\lambda} \geq 0
\end{array}
\end{equation}
These flow rules are identical to those of a Prandtl-Reuss model
(see Eqs. (\ref{flowReuss})). However, they derive from a
different loading function and for this reason the computation of
the plastic multiplier $\dot{\lambda}$ is not the same. When
$f\left( \mbox{\boldmath $\tau$}^{d},R^{d}; \mbox{\boldmath
$\varepsilon$},\mbox{\boldmath $\varepsilon$}^{p},\zeta
\right)=\bar{f}\left(\mbox{\boldmath
$\varepsilon$},\mbox{\boldmath $\varepsilon$}^{p},\zeta
\right)<0$, the loading-unloading conditions reduce to
$\dot{\lambda}=0$, leading to an elastic behavior. As a result,
the function $\bar{f}$ is also called \emph{yielding function,}
while the surface defined by the condition $\bar{f}=0$ is called
\emph{yielding surface}. Conversely, when $\bar{f}\geq 0$ the set
$\mathbb{E}$ evolves by virtue of the dependence of $f$ on the
state variables $\mbox{\boldmath $\varepsilon$},\mbox{\boldmath
$\varepsilon$}^{p}$ and $\zeta$. During this evolution, the actual
thermodynamic forces $(\mbox{\boldmath $\tau$}^{d},R^{d})$ always
satisfy the condition $f=0$. Moreover, the consistency condition
\begin{equation}
\dot{f}=\left[\begin{array}{l} \frac{\partial f}{\partial
\mbox{\boldmath $\tau$}^{d^{\prime}}}:\mbox{\boldmath
$\dot{\tau}$}^{d^{\prime}}+\frac{\partial f}{\partial
R^{d^{\prime}}} \ \dot{R}^{d^{\prime}} \\ \ \ \ \ \ \ \
+\frac{\partial f}{\partial \mbox{\boldmath
$\varepsilon$}}:\mbox{\boldmath
$\dot{\varepsilon}$}+\frac{\partial f}{\partial \mbox{\boldmath
$\varepsilon$}^{p}}:\mbox{\boldmath $\dot{\varepsilon}$}^{p}+
\frac{\partial f}{\partial \zeta} \
\dot{\zeta}\end{array}\right]_{\left(\mbox{\boldmath
$\tau$}^{d^{\prime}}=\mbox{\boldmath
$\tau$}^{d},R^{d^{\prime}}=R^{d}\right)}=0 \label{fdotGenplast}
\end{equation}
is also identically fulfilled and, like for the endochronic
theory, it does not permit to compute $\dot{\lambda}\geq 0$.
Hence, the condition that the so-called \emph{limit function} is
equal to zero has to be invoked and this leads to
\citep{Auricchio95}:
\begin{equation}
\dot{\lambda}=\dot{\zeta}=\left\{
\begin{array}{ll}
0 & \textrm{ \ \ \ if \ \ \ \ }\bar{f}<0 \\
\frac{\left\langle \mathbf{n}:\mbox{\footnotesize\boldmath
$\dot{\varepsilon}$}\right\rangle }{1+ \frac{\bar{N} \left(
\bar{M}-\bar{f}\right) +\left( D_{2}+H_{iso}\right) \ \bar{M}}{2G
\ \bar{f}}} & \textrm{ \ \ \ if \ \ }0\leq \bar{f}\leq \bar{M}
\end{array}
\right.
\end{equation}
where $\bar{M},\bar{N} >0$. It can be proved that when $\bar{f}$
tends to $\bar{M}$, the expression of the plastic multiplier of a
classical plasticity model with linear kinematic and isotropic
hardening is retrieved. Moreover, if $H_{iso}=0$ an asymptotic
value of $\|\mbox{\boldmath $\tau$}^{d}\|$ exists, and is equal to
$\sqrt{\frac{2}{3}}\sigma_{y}+\bar{M}$.

\section{Conclusions}

A common theoretical framework between Prandtl-Reuss models and
endo-chronic theory as well as NLK hardening and generalized
plasticity models was constructed. All models were defined
assuming generalized normality. It was therefore proved that a
unique mathematical structure, based on the notions of
pseudo-potential and generalized normality, was able to contain
plasticity models traditionally formulated by other approaches. In
particular, no extension of the generalized standard class of
materials had to be introduced to describe NLK hardening and
generalized plasticity models. This approach allowed several
comparisons, that have clarified the relationships and analogies
between these, a priori different, plasticity theories.



\clearpage
\begin{appendix}
\section{Appendix}

The vector spaces considered in this paper are: (i) the space of
second order tensors; (ii) the space of \emph{symmetric} second
order tensors $\mathbb{S}^{2}$; (iii) the set of real scalars
$\mathbb{R=}\left( -\infty+\infty\right)$; (iv) the cartesian
product of a finite number of such spaces. They are all equipped
with an Euclidian product, so they are always isomorph to the
Euclidian vector space $\mathbb{X=}$ $\mathbb{R}^{n}.$

\begin{enumerate}
\item A subset $\mathbb{C}$ of $\mathbb{X}$ is said to be:

\begin{description}
\item[(a)] a \emph{convex} set if \ $\left(  1-\lambda\right)  \mathbf{x}
+\lambda\mathbf{y\in}\mathbb{C}$ \ \ whenever \ \ $\mathbf{x,y\in
}\mathbb{C}$\ and\ \ $0<\lambda<1$.

\item[(b)] a \emph{cone} if $\lambda\mathbf{y}\in\mathbb{C}$ when
$\mathbf{y}\in\mathbb{C}$ and $\lambda>0$.
\end{description}

\item Let $\phi:\mathbb{X}\rightarrow(-\infty,\infty]$ be an
extended-real-valued function defined on the vector space
$\mathbb{X}$. Then,
\begin{description}
\item[(a)] the \emph{epigraph} of $\phi$ is the set
\begin{equation}
epi \ \phi=\left\{ \left( \mathbf{y},\mu \right) \textrm{\ such \
that \
}\mathbf{y}\in\mathbb{X},\mu\in\mathbb{R},\mu\geq\phi(\mathbf{y})\right\}
\end{equation}
\item[(b)] $\phi$ is said to be \emph{convex on} $\mathbb{X}$ if $epi \
\phi$ is convex as a subset of $\mathbb{X}\times\mathbb{R}$.
\item[(c)] a convex function $\phi$ is said to be
\emph{proper} if and only if the set
\begin{equation}
\bar{\mathbb{D}}=\left\{  \mathbf{y\in}\mathbb{X}:\phi\left(
\mathbf{y}\right) <+\infty\right\}
\end{equation}
is not empty. The set $\bar{\mathbb{D}}$ is called \emph{effective
domain of $\phi$}, it is convex since $\phi$ is convex and is the
set where $\phi$ is finite.

\item[(d)]  $\phi$ is said to be
\emph{continuous relative to a set} $\bar{\mathbb{D}}$ if the
restriction of $\phi$ to $\bar{\mathbb{D}}$ is a continuous
function.

\item[(e)] $\phi$ is \emph{lower semicontinuous} at
$\mathbf{x\in}\mathbb{X}$ if
\begin{equation}
\phi\left(  \mathbf{x}\right)
=\lim_{\mathbf{y}\rightarrow\mathbf{x}} \ \inf\phi\left(
\mathbf{y}\right)
\end{equation}
It can be proved that the condition of lower semi-continuity of
$\phi$ is equivalent to have that the \emph{level set} $\left\{
\mathbf{y}:\phi\left( \mathbf{y}\right)  \leq\alpha\right\}  $ is
closed in $\mathbb{X}$ for every $\alpha\in\mathbb{R}$ \cite[pg.
51]{Rockafellar69}. As a result, when $\phi$ is a proper convex
function with a (convex) effective domain $\bar{\mathbb{D}}$
\emph{closed} in $\mathbb{X}$ and $\phi$ is continuous relative to
$\bar{\mathbb{D}}$, then $\phi$ is lower-semicontinuous \cite[pg.
52]{Rockafellar69}.
\end{description}

\item Let $\mathbb{X}^{\ast}$ be the dual of $\mathbb{X}$. Since
$\mathbb{X}=\mathbb{R}^{n}$, then
$\mathbb{X}^{\ast}{}^{\ast}=\mathbb{X}$ and the duality product
between $\mathbf{x}$ and $\mathbf{x}^{\ast}$ elements of the dual
vector spaces $\mathbb{X}$ and $\mathbb{X}^{\ast}$ can be written
as $\mathbf{x}^{\ast}\mathbf{\cdot x}$. \newline Let
$\phi:\mathbb{X} \rightarrow(-\infty,\infty]$ be an
extended-real-valued \emph{convex} function. Then, the
\emph{subgradients} of $\phi$ at $\mathbf{x}\in\mathbb{X}$ are
elements $\mathbf{x}^{\ast} \in\mathbb{X}^{\ast}$ such that
\begin{equation}
\forall\mathbf{y}\in\mathbb{X},\quad\phi(\mathbf{y})-\phi(\mathbf{x}
)\geq\mathbf{x}^{\ast}\cdot(\mathbf{y}-\mathbf{x})\label{subgrad}
\end{equation}
The \emph{subdifferential} \emph{set} $\partial\phi(\mathbf{x})$
is the set of all subgradients $\mathbf{x}^{\ast}$ at
$\mathbf{x}$:
\begin{equation}
\partial\phi(\mathbf{x})=\left\{  \mathbf{x}^{\ast}\in\mathbb{X}^{\ast}\textrm{
\ \ such \ that\ the \ condition \ (\ref{subgrad}) \
holds}\right\} \label{subdiff}
\end{equation}
The function $\phi$ is said to be \emph{subdifferentiable} at
$\mathbf{x}$ when $\partial\phi(\mathbf{x})$ is non-empty.

\item If a function $\phi:\mathbb{X}
\rightarrow(-\infty,\infty]$ is \emph{convex, proper,
non-negative} and such that $\phi(\mathbf{0})=0$$,$ then the
\emph{normality} condition
\begin{equation}
\mathbf{x}^{\ast}\in\partial\phi(\mathbf{x})\label{normalx}
\end{equation}
viz. $\mathbf{x}^{\ast}$ belongs to the subdifferential set of
$\phi$ at $\mathbf{x}$, entails that
$\mathbf{x}^{\ast}\mathbf{\cdot x\geq}$ $0$.\newline\emph{Proof:}
Setting $\mathbf{y}=\mathbf{0}$ in the inequality (\ref{subgrad})
entails that, for any $\mathbf{x}$ in the effective domain of
$\phi,$ $-\phi
(\mathbf{x})\geq\mathbf{x}^{\ast}\cdot(\mathbf{0}-\mathbf{x}).$
Hence, by virtue of the non-negativity of $\phi$,
$\mathbf{x}^{\ast}\mathbf{\cdot x}\geq0$.

\item When a function $\phi:\mathbb{X}\rightarrow(-\infty,\infty]$ is proper,
convex and lower semi-continuous, the dual function $\phi^{\ast}
:\mathbb{X}^{\ast}\rightarrow(-\infty,\infty]$ , defined by the
\emph{Legendre-Fenchel transform}
\begin{equation}
\forall\mathbf{y}^{\ast}\in\mathbb{X}^{\ast}\qquad\phi^{\ast}(\mathbf{y}
^{\ast})=\sup_{\mathbf{y}\in\mathbb{X}}(\mathbf{y}^{\ast}\cdot\mathbf{y}
-\phi(\mathbf{y}))\label{fistarconjug}
\end{equation}
is related to $\phi$ by a \emph{one-to-one correspondence}, in the
sense that for such a kind of functions, the conjugate
$\phi^{\ast}$ is in turn proper, convex and lower semi-continuous
and $\phi^{\ast\ast}=\phi$ \cite[pg. 104]{Rockafellar69}.

Under these assumptions, it also holds:
\begin{equation}
\forall\mathbf{y}^{\ast}\in\mathbb{X}^{\ast}\qquad\phi^{\ast}(\mathbf{y}
^{\ast})=\sup_{\mathbf{y}\in\bar{\mathbb{D}}}(\mathbf{y}^{\ast}\cdot\mathbf{y}
-\phi(\mathbf{y}))\label{fistarD}
\end{equation}
Moreover, the following relationships are equivalent:
\newline (i) \ \
$ \mathbf{x}^{\ast} \in \partial\phi(\mathbf{x})$;
\newline  (ii) \
$\mathbf{x}\in\partial\phi^{\ast}(\mathbf{x}^{\ast})$
\newline  (iii)
\vspace{-1cm}
\begin{equation}
\phi(\mathbf{x})+\phi^{\ast}(\mathbf{x}^{\ast})=\mathbf{x}^{\ast}
\cdot\mathbf{x} \label{fenchelEqual}
\end{equation}
Condition (i) is equivalent to $\mathbf{x}^{\ast}\cdot\mathbf{x-}
\phi(\mathbf{x})\geq\mathbf{x}^{\ast}\cdot\mathbf{y-}\phi(\mathbf{y})$.
The supremum of the second term of this inequality is equal by
definition to $\phi^{\ast}(\mathbf{x}^{\ast})$ and occurs when
$\mathbf{y=x}$ and therefore (iii) is the same as (i). Dually,
(ii) and (iii) are equivalent.
\newline \textbf{Remark 1}. Under the previous assumptions, if
$\phi\geq 0$ and $\phi(\mathbf{0})=0$, then (\ref{fistarconjug})
entails that $\phi^{\ast}(\mathbf{0})=0$. Moreover, the identity
$\phi^{\ast\ast}=\phi$ implies that
$\phi(\mathbf{0})=\sup_{\mathbf{y}^{\ast}\in\mathbb{X}^{\ast}}(-\phi^{\ast}(\mathbf{y}^{\ast}))$,
which in turn leads to $\phi^{\ast}\geq 0$. Reciprocally,
$\phi^{\ast} \geq 0$ and $\phi^{\ast}(\mathbf{0})=0$ entail that
$\phi\geq 0$ and $\phi(\mathbf{0})=0$.
\newline \textbf{Remark 2}. If $\phi^{\ast}$
is such that $\phi^{\ast} \geq 0$ and $\phi^{\ast}(\mathbf{0})=0$,
then the normality condition (ii) implies that
$\mathbf{x}^{\ast}\cdot\mathbf{x} \geq 0$. \emph{Proof:} Condition
(ii) is equivalent to (i), with $\phi\geq 0$ and
$\phi(\mathbf{0})=0$. Then, using the result of item 4, the
non-negativity of $\mathbf{x}^{\ast}\cdot\mathbf{x}$ follows.
\newline \textbf{Remark 3}. The conjugate $\tilde{\phi}^{\ast}$ of an arbitrary function
$\tilde{\phi}:\mathbb{X}\rightarrow(-\infty,\infty]$ can still be
defined by (\ref{fistarconjug}). In this case,
$\tilde{\phi}^{\ast}$ is proper, convex, lower semi-continuous and
is equal to the conjugated $\phi^{\ast}$ of $\phi=cl\left( conv\
\tilde{\phi}\right)$, where $\phi$ is the greatest proper convex
lower semi-continuous function majorized by $\tilde{\phi}$
\cite[pp. 52, 103-104]{Rockafellar69}.

\item A function $\phi :\mathbb{X}\rightarrow (-\infty ,\infty ]$
is \emph{positively homogeneous of order 1} if and only if
\begin{equation}
\forall \mathbf{y}\in\mathbb{X}, \
\forall\rho\in(0,\infty)\textrm{, \ \ \ \ }\phi \left( \rho
\mathbf{y}\right) =\rho \phi \left( \mathbf{y}\right)
\end{equation}
The epigraph of such functions is a cone \citep[pg. 30
]{Rockafellar69}.

Given $\phi :\mathbb{X}\rightarrow (-\infty ,\infty ]$, the
following three statements are equivalent:

\begin{description}
\item[(i)] $\phi $ is proper, convex, lower semi-continuous and positively
homogeneous of order 1

\item[(ii)] The Legendre-Fenchel conjugate $\phi ^{\ast }$ of $\phi $ is the
indicator function of a non-empty, convex and closed set
$\bar{\mathbb{E}}$, i.e.
\[
\phi ^{\ast }\left( \mathbf{y}^{\ast }\right)
=\mathbb{I}_{\bar{\mathbb{E}}}\left( \mathbf{y}^{\ast }\right)
=\left\{
\begin{array}{cc}
0 & \textrm{ \ if }\mathbf{y}^{\ast }\in \bar{\mathbb{E}} \\
+\infty  & \textrm{ \ if }\mathbf{y}^{\ast }\in \bar{\mathbb{E}}
\end{array}
\right.
\]

\item[(iii)] $\phi $ is the \emph{support function} of a non-empty, convex and
closed set $\bar{\mathbb{E}}$, i.e.
\[
\phi \left( \mathbf{y}\right) =\mathbb{I}_{\bar{\mathbb{E}}}^{\ast
}\left( \mathbf{y}\right) =\sup_{\mathbf{y}^{\ast }\in
\bar{\mathbb{E}}}\left( \mathbf{y}^{\ast }\mathbf{\cdot y}\right)
\]
\end{description}

The equivalence between (i) and (ii) can be proved by showing that
$\phi ^{\ast }$ has no values other than $0$ and $+\infty $
\citep[pg. 114]{Rockafellar69}. The set where $\phi ^{\ast }=0$ is
non-empty, convex and closed since $\phi$ is proper, convex and
lower semi-continuous. The equivalence between (ii) and (iii)
follows from the definition of Legendre-Fenchel transform, support
functions and indicator functions.

\textbf{Remark}. If $\phi $ fulfils conditions in (i), then for
any $\mathbf{x}$ where $\phi $ is subdifferentiable,
\[
\phi \left( \mathbf{x}\right)=\phi^{\ast\ast} \left(
\mathbf{x}\right) =\mathbf{x}^{\ast }\mathbf{\cdot x} \textrm{\ \
\ \ with \ \ }\mathbf{x}^{\ast}\in\partial\phi(\mathbf{x})
\]\emph{Proof:} from the equivalence between (i) and (ii),
the conjugated of $\phi$ is the indicator function of a closed
convex set $\bar{\mathbb{E}}$ and $\mathbf{x}^{\ast }\in
\bar{\mathbb{E}}$ since $\phi $ is subdifferentiable at
$\mathbf{x}$ by assumption. Then, use Eq. (\ref{fenchelEqual}) and
recall by (ii) that $\phi ^{\ast }\left( \mathbf{x}^{\ast }\right)
=0$.

\item Let $\phi:\mathbb{X}\rightarrow(-\infty,+\infty]$ be a
proper, convex, lower semi-continuous function, positively
homogeneous of order 1. Then: \begin{description}
\item [(i)] from item (6), its conjugate
$\phi^{\ast}$ is the indicator function of a non-empty, closed and
convex set $\bar{\mathbb{E}}$. Hence, by using the definition
(\ref{subgrad}),
\begin{equation}
\partial\phi^{\ast}\left(  \mathbf{x}^{\ast}\right)
=\partial\mathbb{I}_{\bar{\mathbb{E}}}\left(
\mathbf{x}^{\ast}\right) =\left\{
\begin{array}
[c]{ll} \mathbf{0} & \textrm{ \ \ \ if \ \ }\mathbf{x}^{\ast}\in
int\left(
\bar{\mathbb{E}}\right)  \\
\mathcal{C}\left(  \mathbf{x}^{\ast}\right)   & \textrm{ \ \ \ if
\ \ }
\mathbf{x}^{\ast}\in\partial\bar{\mathbb{E}}\\
\varnothing & \textrm{ \ \ \ if \ \
}\mathbf{x}^{\ast}\notin\bar{\mathbb{E}}
\end{array}
\right.  \label{SubDiffGen}
\end{equation}
where $\mathcal{C}\left(  \mathbf{x}^{\ast}\right)  =\left\{
\mathbf{x} \in\mathbb{X}\ \ :\
\forall\mathbf{y}^{\ast}\in\bar{\mathbb{E}\quad}\mathbf{x\cdot
}\left( \mathbf{y}^{\ast}-\mathbf{x}^{\ast}\right)  \leq0\right\}
$ is the so-called \emph{normal cone} at $\ \mathbf{x}^{\ast}\in
\partial\bar{\mathbb{E}}$.
\item [(ii)] if in addition $\phi$
\emph{does not depend on some components $\mathbf{y}_{1}$ of}
$\mathbf{y}=(\mathbf{y}_{1},\mathbf{y}_{2})\subset\mathbb{X=X}_{1}
\times\mathbb{X}_{2}$ , i.e. $\phi(\mathbf{y})=\phi(\mathbf{y}_{1}
,\mathbf{y}_{2})=\hat{\phi}(\mathbf{y}_{2}),$ then the conjugated
function $\phi^{\ast}$ can be computed as follows:
\begin{equation}
\begin{array}[c]{l}
\phi^{\ast}(\mathbf{y}_{1}^{\ast},\mathbf{y}_{2}^{\ast})=\sup_{\left(
\mathbf{y}_{1},\mathbf{y}_{2}\right)
\in\mathbb{X}}(\mathbf{y}_{1}^{\ast
}\cdot\mathbf{y}_{1}+\mathbf{y}_{2}^{\ast}\cdot\mathbf{y}_{2}-\hat{\phi
}(\mathbf{y}_{2}))\\
\ \ \ \ \ \ \ \ \ \ \ \ \
=\mathbb{I}_{0}(\mathbf{y}_{1}^{\ast})+\sup
_{\mathbf{y}_{2}\in\mathbb{X}_{2}}(\mathbf{y}_{2}^{\ast}\cdot\mathbf{y}
_{2}-\hat{\phi}(\mathbf{y}_{2}))\\
\ \ \ \ \ \ \ \ \ \ \ \ \
=\mathbb{I}_{0}(\mathbf{y}_{1}^{\ast})+\mathbb{I}_\mathbb{E}
(\mathbf{y}_{2}^{\ast})
\end{array}
\end{equation}
The Legendre-Fenchel conjugate is the indicator function of
$\mathbf{0}$ with respect to $\mathbf{y}_{1}^{\ast}$ plus the
Legendre-Fenchel conjugate of $\hat{\phi}(\mathbf{x}_{2})$, which
is the indicator function of a non-empty, closed and convex set
$\mathbb{E}$. Hence,
\[
\mathbf{x}\in\partial\mathbb{I}_{\bar{\mathbb{E}}} \left(
\mathbf{x}^{\ast}\right) \ \ \ \Leftrightarrow \ \ \
\mathbf{x}_{1}\in \mathbb{X}_{1} \ \ \textrm{and} \ \
\mathbf{x}_{2}\in\partial\mathbb{I}_\mathbb{E}(\mathbf{x}
_{2}^{\ast})
\]
In the particular case where $\mathbb{E} \mathbf{=}\left\{
\mathbf{y}^{\ast}\in\mathbb{X}^{\ast}\ \textrm{ \ such \ that \ }\
f\left( \mathbf{y}^{\ast}\right)  \leq0\right\}  ,$ where $f$ is a
convex and smooth function, the normality condition at
$\mathbf{y^{\ast}}=\mathbf{x^{\ast}} $, viz.
$\mathbf{x\in\partial}\mathbb{I}_{\mathbb{E}}\left(
\mathbf{x}^{\ast}\right)$, can be written as follows
\[
\mathbf{x}=\mu \ grad \ f\left( \mathbf{x}^{\ast }\right) \textrm{
\ \ \ \ \ \ with \ \ \ \ }\left\{
\begin{array}{l}
\mu =0\textrm{ \ \ \ \ for \  }f\left( \mathbf{x}^{\ast }\right) <0 \\
\mu \geq 0\textrm{ \ \ \ \ for \ }f\left( \mathbf{x}^{\ast
}\right) =0
\end{array}
\right.
\]
These two last conditions are often replaced by
\begin{equation}
\mu\geq0,\ \ \ f\left( \mathbf{x}^{\ast }\right)\leq0,\ \ \ \ \mu
f\left( \mathbf{x}^{\ast }\right)=0 \label{kuhntucker}
\end{equation}
which are the classical loading-unloading conditions of
plasticity, usually written with $\mu$ replaced by the
\emph{plastic multiplier} $\dot{\lambda}$. The dependence of $f$
on the argument $\mathbf{x}^{\ast }$ is often omitted in order to
simplify the notation. In the convex mathematical programming
literature, (\ref{kuhntucker}) are known as Kuhn-Tucker conditions
(see e.g. \citet{Luenberger84}).
\end{description}

\end{enumerate}

\end{appendix}


\clearpage
\begin{figure}
\begin{center}
\includegraphics[width=13.4cm]{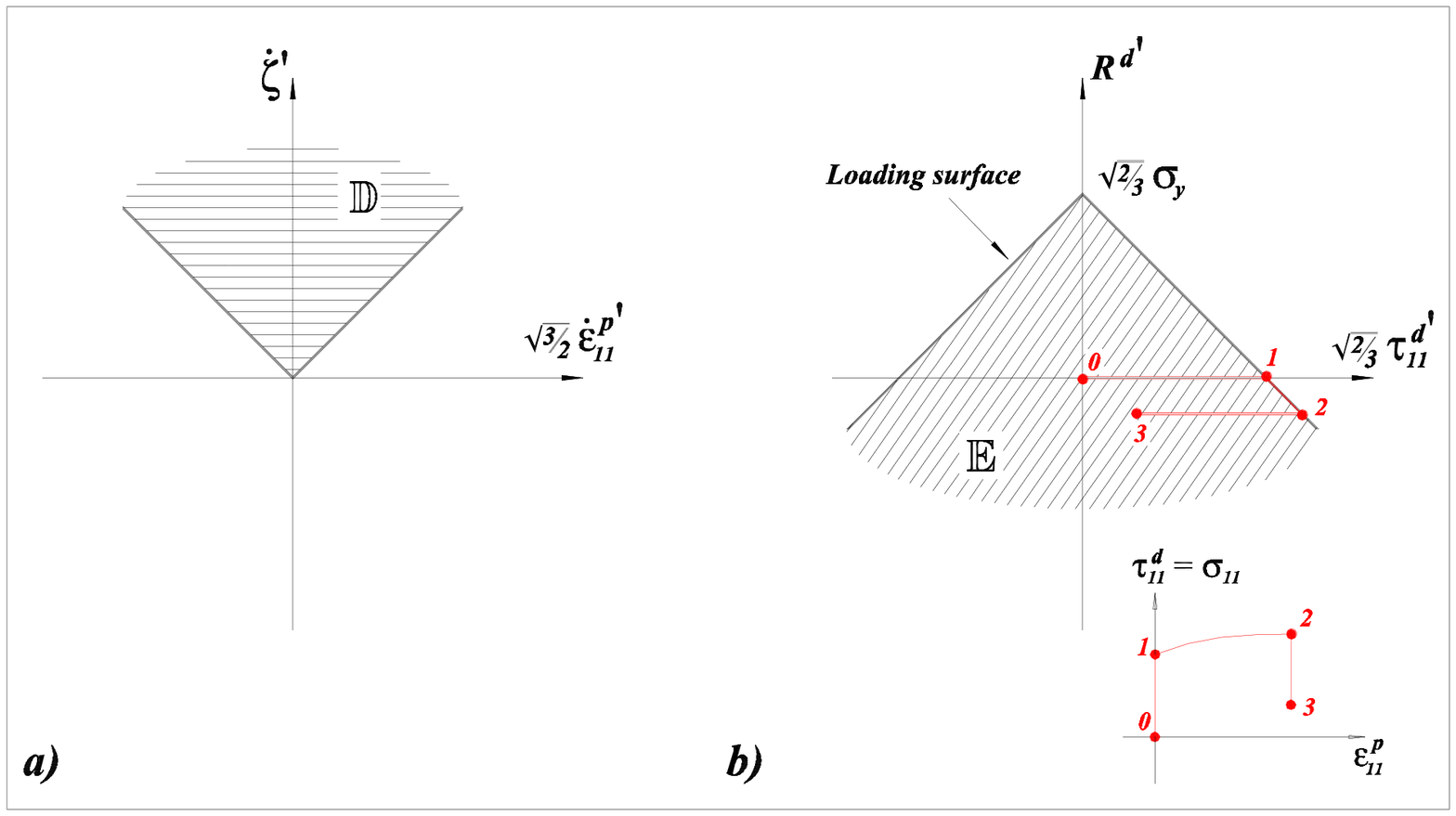}
\caption{Classical Prandtl-Reuss model. Tension-compression case.
\newline a) Projection of the pseudo-potential effective domain
$\bar{\mathbb{D}}$ on the $\left(\mbox{\boldmath
$\dot{\varepsilon}$}^{p^{\prime}},\dot{\zeta}^{\prime}\right)$-plane.
This set is indicated by $\mathbb{D}$. b) Domain $ \mathbb{E}$
associated with the dual pseudo-potential $\phi^{*}$.}

\label{FigPrandtl1}
\end{center}
\end{figure}

\clearpage

\clearpage

\clearpage

\clearpage

\clearpage

\clearpage

\clearpage

\clearpage

\clearpage \textbf{Silvano Erlicher} received the Dip.-Eng. and
Ph.D. degrees in Civil Engineering from the University of Trento,
Italy, in 1999 and 2003, respectively. From May 2003 to April
2004, he held a post-doctoral research position at ENPC-LCPC,
Paris, France. In 2004, he joined the Laboratoire d'Analyse des
Mat\'{e}riaux et Structures (LAMI-ENPC/Institut Navier), Paris,
France. His research interests include seismic engineering with
special emphasis on cyclic inelastic behavior of materials and
structures.

\textbf{Nelly Point} obtained the \emph{agr\'{e}gation} in
Mathematics as student at the Ecole Normale Sup\'{e}rieure (ENS)
in Paris, France. She received her Ph.D. degree in 1975 and her
\emph{th\`{e}se d'\'{e}tat} in 1989. She joined the Conservatoire
National des Art et M\'{e}tiers (CNAM) in 1973, where she teaches
Mathematics for engineers. She develops her research activity at
the Laboratoire d'Analyse des Mat\'{e}riaux et Identification
(LAMI-ENPC/Institut Navier), Paris, France. Her research interests
include applied mathematics and mechanics, in particular adhesion,
fracture, plasticity and hysteresis.

\end{document}